\documentclass[11pt,a4paper]{article}
 \pdfoutput=1
\usepackage{jcappub}

\usepackage{tabularx}
\usepackage{subcaption}
\usepackage{caption}
\usepackage{enumerate}
\usepackage{amsmath,amssymb,mathtools}
\usepackage{graphicx}
\usepackage{dcolumn}
\usepackage{bm,color}
\usepackage{hyperref}
\usepackage{accents}
\usepackage{amssymb,float}
\usepackage{amsmath}
\usepackage{multirow}
\usepackage{pifont}
\usepackage{tikz}
\usetikzlibrary{positioning}
\def\layersep{2.5cm}

\hypersetup{
    colorlinks=true,
    linkcolor=red,
    filecolor=magenta,      
    citecolor=blue
}

\title{Neural Network Reconstruction of Late--Time Cosmology and Null Tests}

\author[a,b]{Konstantinos Dialektopoulos,}
\author[c,d]{Jackson Levi Said,}
\author[c]{Jurgen Mifsud,}
\author[e]{Joseph Sultana,}
\author[c,d]{and Kristian Zarb Adami}

\affiliation[a]{Laboratory of Physics, Faculty of Engineering, Aristotle University of Thessaloniki, Thessaloniki 54124, Greece}
\affiliation[b]{Center for Gravitation and Cosmology, College of Physical Science and Technology, Yangzhou University, Yangzhou 225009, China}
\affiliation[c]{Institute of Space Sciences and Astronomy, University of Malta, Malta, MSD 2080}
\affiliation[d]{Department of Physics, University of Malta, Malta, MSD 2080}
\affiliation[e]{Department of Mathematics, University of Malta, Msida, MSD 2080, Malta}

\emailAdd{kdialekt@gmail.com}
\emailAdd{jackson.said@um.edu.mt}
\emailAdd{jurgen.mifsud@um.edu.mt}
\emailAdd{joseph.sultana@um.edu.mt}
\emailAdd{kristian.zarb-adami@um.edu.mt}

\date{\today}

\abstract{
The prospect of nonparametric reconstructions of cosmological parameters from observational data sets has been a popular topic in the literature for a number of years. This has mainly taken the form of a technique based on Gaussian processes but this approach is exposed to several foundational issues ranging from overfitting to kernel consistency problems. In this work, we explore the possibility of using artificial neural networks (ANN) to reconstruct late--time expansion and large scale structure cosmological parameters. We first show how mock data can be used to design an optimal ANN for both parameters, which we then use with real data to infer their respective redshift profiles. We further consider cosmological null tests with the reconstructed data in order to confirm the validity of the concordance model of cosmology, in which we observe a mild deviation with cosmic growth data.
}

\begin{document}

\maketitle
\flushbottom

\section{\label{sec:intro}Introduction}

Late--time cosmic acceleration is best described in general relativity (GR) by the inclusion of cold dark matter (CDM) and a cosmological constant ($\Lambda$) which are both embodied in the $\Lambda$CDM concordance model \cite{Peebles:2002gy,Copeland:2006wr}. In this setting, CDM acts as a stabilizer for galactic structures \cite{Baudis:2016qwx,Bertone:2004pz}, while on larger scales the cosmological constant takes on the role of dark energy and is responsible for the late--time accelerated expansion of the Universe \cite{Riess:1998cb,Perlmutter:1998np}. Putting aside internal consistency issues \cite{RevModPhys.61.1}, this description of the Universe is increasingly coming into tension with observations \cite{DiValentino:2020vhf,DiValentino:2020zio,DiValentino:2020vvd,Staicova:2021ajb}, while the prospect of detecting CDM seems to be ever more elusive \cite{Gaitskell:2004gd}.

This has led to the consideration of a host of new physical models through which current and future observation can be more precisely described. These new proposals range from a reconsideration of GR as the fundamental theory of gravitation \cite{Clifton:2011jh,CANTATA:2021ktz,Bahamonde:2021gfp,AlvesBatista:2021gzc,Addazi:2021xuf} to new aspects of CDM \cite{Feng:2010gw,Dodelson:1993je} as well as dynamical dark energy models \cite{Copeland:2006wr,Benisty:2021gde,Benisty:2020otr}, among others. For this reason, it is crucial that observational data sets can be used to produce cosmological parameters without a dependence on a cosmological model, such as $\Lambda$CDM. This can become more intricate as more complex parameters are considered such as the growth of large scale structure data. To this end, a whole new range of statistical approaches have been utilized to build cosmological parameter profiles without the need of an underlying physical model. These model-independent techniques are nonparametric in that they do not necessitate a physical model to interpret the data, but they do require assumptions in the statistical interpretation of each parameter data set.

Reconstruction techniques that are model independent have since become more robust and better understood in terms of late--time cosmological data sets. The most popular of these methods is Gaussian processes (GP) \cite{RasmussenW06} which has been successfully used to produce the Hubble diagram for various data sets \cite{Busti:2014aoa,Gomez-Valent:2018hwc,Briffa:2020qli} as well as large scale structure growth data \cite{Benisty:2020kdt,LeviSaid:2021yat}. GP assumes every element of a data set is normally distributed and part of a larger stochastic process, by optimizing a covariance function between these points it can reconstruct the entire evolution of the data set for some ranges of the data. The immediate issue here is that not all cosmological data is normally distributed. More concretely there are auxiliary issues with selecting the covariance function, but this can be circumvented with genetic algorithms \cite{Bernardo:2021mfs} to a certain extent. However, GP suffers from a bigger issue with over-fitting for low redshift data \cite{OColgain:2021pyh}, which impacts inferred values of $H_0$ in the Hubble diagram, and $f\sigma_{8_0}$ for growth data. These challenges of GP seem to be generically problematic features that are hard to avoid within this approach. On the other hand, other approaches to reconstruction exist such as the Locally weighted Scatterplot Smoothing together with Simulation and extrapolation method (LOESS-Simex) \cite{Montiel:2014fpa,Escamilla-Rivera:2015odt}. LOESS-Simex provides an independent nonparametric approach through which it generalizes the technique of least-squares in reconstructing data sets. Both GP and LOESS have been compared against each other in Ref.~\cite{Escamilla-Rivera:2021rbe} where their performance in reconstructing data sets was quantified using a number of statistical metrics.

Another nonparametric approach that can be used to reconstruct late--time cosmology data is an artificial neural network (ANN). Inspired from biological neural networks, a network is a setup composed of a collection of neurons which are organized into layers \cite{aggarwal2018neural,Wang:2020sxl,Gomez-Vargas:2021zyl}. Each neuron is built on a so-called activation function which governs the neuron output. Thus, every neuron contains a number of hyperparameters (statistical parameters) which must be assigned a value. To do this the network is trained with observational data which optimize the neural responses. This then produces an ANN that can mimic the observational data at every redshift. The core difference between GP and ANNs is that GP entails a supervised form of learning, meaning that the data is used concurrently as the reconstruction takes place, while ANNs are unsupervised, which means that once the neural network is trained it no longer requires the data to reconstruct a parameter at new redshifts since it is imitating the natural process itself.

There have been a number of works on using GP to reconstruct the Hubble diagram using a variety of data sets as reported in Refs.~\cite{Qi:2016wwb,Lin:2019cuy,Singirikonda:2020ieg,Bengaly:2020neu,Velasquez-Toribio:2021ufm,Reyes:2021owe,vonMarttens:2018bvz,vonMarttens:2020apn,Andrade:2021njl}. In light of the current disagreement between the inferred values of the Hubble constant $(H_0)$, primarily between the local and early--time determinations of $H_0$, we shall be analysing the impacts of $H_0$ priors on our ANN reconstructions of the cosmic expansion in this work. Several local measurements of $H_0$ have been reported (see, for instance, Refs. \cite{Blakeslee:2021rqi,Kourkchi:2020iyz,Schombert:2020pxm,LIGOScientific:2017adf,LIGOScientific:2019zcs,Mukherjee:2019qmm,DES:2019fny,Birrer:2020tax,Pesce:2020xfe,Khetan:2020hmh,Wong:2019kwg,Huang:2019yhh,Riess:2019cxk,Riess:2020fzl,Freedman:2020dne,Freedman:2019jwv,Denzel:2020zuq}), where one could easily observe that the reported late--time measurements seem to agree on $H_0\gtrsim70\,\mathrm{km}\,\mathrm{s}^{-1}\mathrm{Mpc}^{-1}$, at the
very least. We will therefore be considering the most precise Cepheid calibration result of $H_0^\mathrm{R20}=73.2\pm1.3\,\mathrm{km}\,\mathrm{s}^{-1}\mathrm{Mpc}^{-1}$ \cite{Riess:2020fzl} (R20) along with $H_0^\mathrm{TRGB}=69.8\pm1.88\,\mathrm{km}\,\mathrm{s}^{-1}\mathrm{Mpc}^{-1}$ \cite{Freedman:2020dne,Freedman:2019jwv} which has been recently inferred via the Tip of the Red Giant Branch (TRGB) calibration technique. The discrepancy between early--time and late--time measurements of $H_0$ currently stands at the level of $\sim4-6\sigma$ tension \cite{Verde:2019ivm}, with R20 being the most discordant measurement. It is well--known that the most precise early--time determination of $H_0=67.36\pm0.54\,\mathrm{km}\,\mathrm{s}^{-1}\mathrm{Mpc}^{-1}$ \cite{Aghanim:2018eyx} is dependent on the adopted cosmological model, since we are probing the physics of the cosmic microwave background (CMB) with which we can indirectly infer the current expansion rate of the Universe. 

Several studies tried to address such a puzzling discrepancy in a number of avenues, such as by the introduction of new physics and via artifacts of systematic errors \cite{Efstathiou:2021ocp,Mortsell:2021nzg,Mortsell:2021tcx,Freedman:2021ahq}. For instance, Ref. \cite{Beenakker:2021vff} outlines seven key theoretical assumptions which might be broken to alleviate this tension, while modified cosmological frameworks and exotic physics proposals have been put forward to explain such a discrepancy (see, for instance, Refs. \cite{Nunes:2018xbm,Poulin:2018zxs,Zhou:2021xov,DeFelice:2020sdq,DeFelice:2020cpt,Thiele:2021okz,Dainotti:2021pqg,Banihashemi:2018oxo,Guo:2018ans,Poulin:2018cxd,Kreisch:2019yzn,Vattis:2019efj,Lin:2019qug,DiValentino:2019exe,Vagnozzi:2019ezj,Abadi:2020hbr,DiValentino:2019ffd,Vagnozzi:2021gjh,Ye:2020btb,Akarsu:2019hmw,Gonzalez:2020fdy,Vagnozzi:2021tjv,Blinov:2021mdk,vanPutten:2021hlu,DiValentino:2020vnx,Okamatsu:2021jil,Hill:2021yec,delaMacorra:2021hoh,Freese:2021rjq,Wang:2021kxc,Gurzadyan:2021jrw,Huang:2021dba,Braglia:2020auw,Bernardo:2021qhu}, and Refs. \cite{DiValentino:2021izs,Shah:2021onj} for recent reviews). A considerable number of models tried to alleviate the $H_0$ tension from the point of view of the CMB calibration technique which is derived directly from the measurement of the angular scale subtended by the sound horizon scale $r_s$. One should remark that all the available probes that adopt the value of $r_s$ within the concordance model of cosmology are characterised by relatively low values of $H_0$, and this was therefore the main driving force behind the proposal of theoretical frameworks which modify the physics prior to last scattering. However, Ref. \cite{Jedamzik:2020zmd} showed that the modification of $r_s$ only might not be enough to address the $H_0$ tension completely, and we are therefore still lacking a compelling model addressing the Hubble tension while keeping a good fit to all available data. Moreover, Ref. \cite{Baxter:2020qlr} considered the possibility of measuring $H_0$ from the CMB lensing power spectrum in a way that is independent of $r_s$. Indeed, they reported a slightly higher value of $H_0=73.5\pm5.3\,\mathrm{km}\,\mathrm{s}^{-1}\mathrm{Mpc}^{-1}$ which is more consistent with late--time measurements of the Hubble constant, although the inferred errors with current data sets are relatively large which should improve with upcoming cosmological surveys.

Another important cosmological parameter that is increasingly coming into tension is the growth of large scale structure where early--time model dependent measurements appear to be in tension with late--time observations \cite{DiValentino:2020vvd}. We consider this through the $f\sigma_8(z)$ parameter which quantifies the growth rate of cosmological perturbations and matter power spectrum overdensities normalized on scales of $8 h^{-1} {\rm Mpc}$ \cite{Kazantzidis:2018rnb}. This tension has been reported at the level of $\sim2 - 3.5\sigma$ with respect to Planck estimates, and even higher in some studies \cite{Heymans:2020gsg,Hildebrandt:2016iqg,Kuijken:2015vca}. For this reason we also consider this cosmological parameter in our work, since it is timely to consider a data driven reconstruction of the currently available $f\sigma_8(z)$ data points, which would provide an alternative avenue to understand this reported tension instead of solely focusing on the proposal of alternative models which would normally introduce extra degrees of freedom (see, for instance, Refs. \cite{Barros:2020bgg,Zumalacarregui:2020cjh,Choi:2020pyy,Heimersheim:2020aoc,Chamings:2019kcl,DiValentino:2019jae,DiValentino:2019ffd,Buen-Abad:2017gxg,Abellan:2021bpx,DES:2020mpv,Abellan:2020pmw,Xiao:2019ccl,Pandey:2019plg,Chudaykin:2017ptd,DiValentino:2017oaw,Enqvist:2015ara,Feng:2017nss,Mccarthy:2017yqf,MacCrann:2014wfa} and references therein). The growth rate of matter density perturbations $f(z)$, is inferred from the peculiar velocities arising from redshift space distortions (RSD) \cite{Kaiser:1987qv} measurements in galaxy redshift surveys which are a velocity--induced mapping from real--space to redshift--space due to the line--of--sight peculiar motions of objects that introduce anisotropies in their clustering patterns. Such an effect depends on the growth of cosmic structure, hence making RSD probes sensitive to the combination of $f\sigma_8(z)$. 

From the ANN reconstruction results of the cosmic expansion as well as from the cosmic growth history, we will be quantifying the deviations of observational data from the concordance model of cosmology via two formulations of null tests. Rather than a parametric analysis which is carried out via a model--fitting technique of alternative cosmological models, the considered null tests will be based on general consistency relations that the concordance model of cosmology must obey. These consistency relations are also easy to be interpreted since these are constant if the Universe is described by the $\Lambda$CDM model regardless of the parameters of the model. In this way, we can determine how well we will be able to rule out the concordance model without prior assumptions on the underlying cosmological parameters. Such a nonparametric analysis avoids biasing the derived results by fitting specific cosmological models, which makes our null test analysis as model--independent as possible. 

The motivation for this work is to further advance the prospect of model-independent reconstruction techniques for late--time cosmological data using ANNs. We do this by first providing more information about ANNs in our context in Sec.~\ref{sec:method}. We immediately apply this to expansion data in Sec.~\ref{sec:hubble} where we describe how we train our ANN. We perform a similar analysis for $f\sigma_8$ data in Sec.~\ref{sec:fs8} but with some caveats. Finally, we close with a summary and conclusion in Sec.~\ref{sec:conc}.

\section{\label{sec:method}Methodology}

\tikzset{%
  every neuron/.style={
    circle,
    fill=green!70,
    minimum size=32pt, inner sep=0pt
  },
  mid neuron/.style={
    circle,
    fill=blue!70,
    minimum size=32pt, inner sep=0pt
  },
  last neuron/.style={
    circle,
    fill=red!70,
    minimum size=32pt, inner sep=0pt
  },
  neuron missing/.style={
    draw=none,
    fill=none,
    scale=4,
    text height=0.333cm,
    execute at begin node=\color{black}$\vdots$
  },
}
\begin{figure}[t!]
    \centering
    \begin{tikzpicture}[shorten >=1pt,->,draw=black!50, node distance=\layersep]
    \tikzstyle{annot} = [text width=5em, text centered]

\foreach \m/\l [count=\y] in {1}
  \node [every neuron/.try, neuron \m/.try] (input-\m) at (0,-1.5*\y) {};

\foreach \m [count=\y] in {1,2,3,missing,4}
  \node [mid neuron/.try, neuron \m/.try ] (hidden-\m) at (5,2-\y*1.5) {};

\foreach \m [count=\y] in {1,2}
  \node [last neuron/.try, neuron \m/.try ] (output-\m) at (10,1.25-2*\y) {};

\foreach \name / \y in {1}
    \path[yshift=-.1cm] node[above] (input+\name) at (0,-1.6\name) {\large$z$};

\foreach \l [count=\i] in {1,2,3,k}
  \node[below] at (hidden-\i) {\large$\mathfrak{n}_\l$};

\foreach \name / \y in {{\large $\Upsilon(z)$} / 1, {\large$\sigma_\Upsilon^{}(z)$} / 2}
        \path[yshift=-.1cm] node[above, right of=H-3] 
        (output-\y) at (7.5,1.35-2*\y) {\name};

\foreach \i in {1}
  \foreach \j in {1,2,3,...,4}
    \draw [->] (input-\i) -- (hidden-\j);

\foreach \i in {1,2,3,...,4}
  \foreach \j in {1,2}
    \draw [->] (hidden-\i) -- (output-\j);

\foreach \l [count=\x from 0] in {\large Input, \large Hidden, \large Output}
  \node [align=center, above] at (\x*5,2) {\l \\ \large layer};

\end{tikzpicture}
    \caption{The general structure of the adopted ANN, where the input is the redshift of a cosmological parameter $\Upsilon(z)$, and the outputs are the corresponding value and error of $\Upsilon(z)$.}
\label{fig:ANN_structure}
\end{figure}
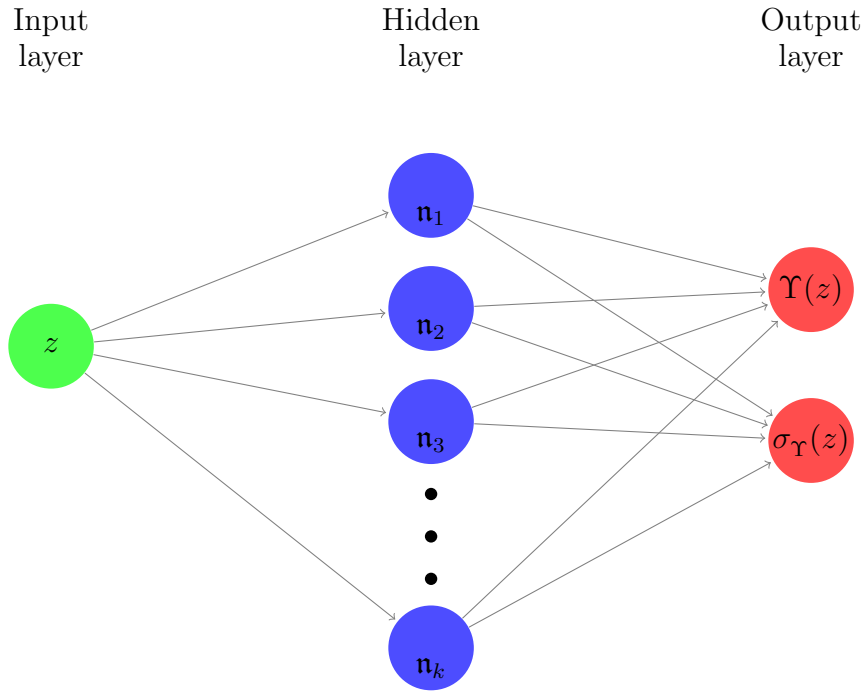

We will now briefly outline the adopted ANN technique \cite{2015arXiv151107289C} which will be used in the reconstruction of the cosmic expansion data in Sec.~\ref{sec:hubble}, as well as the reconstruction of the cosmological growth rate in Sec.~\ref{sec:fs8}. The general structure of an ANN is composed of an input layer that is connected to a hidden layer, or a series of successive hidden layers, and an output layer, where the elements of each layer are known as neurons. The first layer consists of the features of the data set, which will consequently provide the values for the first layer of neurons. On the other hand, the output layer consists of neurons whose values must be evaluated by an error function which measures the difference between the value given by the ANN and the expected one. We should remark that these input, hidden and output layers give structure to the network and provide a direction that input signals must take for the network output. We illustrate such a structure for a one hidden layer ANN with $\mathfrak{n}_k$ neurons in Fig. \ref{fig:ANN_structure}, such that for each redshift point at the input layer, it outputs a generic cosmological parameter $\Upsilon(z)$ and its corresponding uncertainty $\sigma_\Upsilon^{}(z)$. 

The ANN process involves the application of a linear transformation (composed of linear weights and biases) and a nonlinear activation on the input layer, and then the inferred results are propagated to the succeeding layer, until a linear transformation is applied to the output layer. In this way, an input signal will traverse the entire network in a structured manner.

With the use of activation functions, which modify the data they receive before transmitting it to the next layer, an ANN is able to model highly complex relationships between features encapsulated in the data. We will be considering the Exponential Linear Unit (ELU) \cite{2015arXiv151107289C} as the activation function, specified by
\begin{equation}
    f(x) = 
  \begin{cases} 
   {x} & \text{if } x>0 \\
   {\alpha(e^x-1)} & \text{if } x \leq 0
  \end{cases}\,,
\end{equation}
where $\alpha$ is a positive hyperparameter that controls the value to which an ELU saturates for negative net inputs, which we set to unity. There is a range of standard activation functions with the ELU being a very popular one. We should remark that an ANN model is characterised by several intrinsic parameters, better known as hyperparameters, such as the number of layers, number of neurons, optimiser algorithm, among others.

As already mentioned, in order to optimise the parameters of an ANN, the difference between the predicted result $\hat{\mathcal{Y}}$ and the ground truth $\mathcal{Y}$ is minimised during the training process of the ANN, which is then quantitatively mapped to a specific loss function. The loss function is minimised by an optimisation algorithm such as gradient descent combined with the back--propagation algorithm to calculate gradients. We shall be considering the mean absolute error loss function in the following analyses, better known as the L1 loss function, which minimises the absolute differences between $\hat{\mathcal{Y}}$ and $\mathcal{Y}$. However, we also consider the ANN reconstructions with the mean squared error (MSE) loss function that minimises the squared differences between $\hat{\mathcal{Y}}$ and $\mathcal{Y}$, along with the smooth L1 (SL1) loss function which uses a squared term if the absolute error falls below unity and absolute term otherwise. These loss functions quantify the degree to which the input data is modeled by the output reconstruction. The network parameters are updated by a gradient--based optimiser in each iteration. In our work, we adopt Adam's algorithm \cite{2014arXiv1412.6980K} as our optimiser, since this has also accelerated the convergence. 

It should be noted that an ANN with at least one hidden layer with a finite number of neurons can approach any continuous function if the activation function is continuous and non--linear \cite{HORNIK1990551}, therefore ANNs are applicable to cosmological data sets as they fulfil the latter requirements. We shall be using the \texttt{PyTorch}\footnote{\url{https://pytorch.org/docs/master/index.html}} based code for reconstructing functions from data called Reconstruct Functions with ANN (\texttt{ReFANN}\footnote{\url{https://github.com/Guo-Jian-Wang/refann}}) \cite{Wang:2019vxv}. Although this code could be used on CPUs, we ran this code on GPUs due to a significant decrease in computational time. We further make use of batch normalisation \cite{2015arXiv150203167I} which is implemented prior to every nonlinear layer, and which also accelerates the convergence due to the stabilisation in the distribution among ANN variables.

\section{\label{sec:hubble}Hubble data}

We first consider the generation of the $H(z)$ mock data points and the ANN training process in Sec. \ref{sec:hubble_training}, which will be used for structuring the number of layers and neurons of the ANN. The ANN will be used to reconstruct the Hubble diagram and perform a null test of the concordance model of cosmology.

\subsection{\label{sec:hubble_training}Simulation and training of Hubble data}

\begin{figure}
    \centering
    \includegraphics[width=0.485\linewidth]{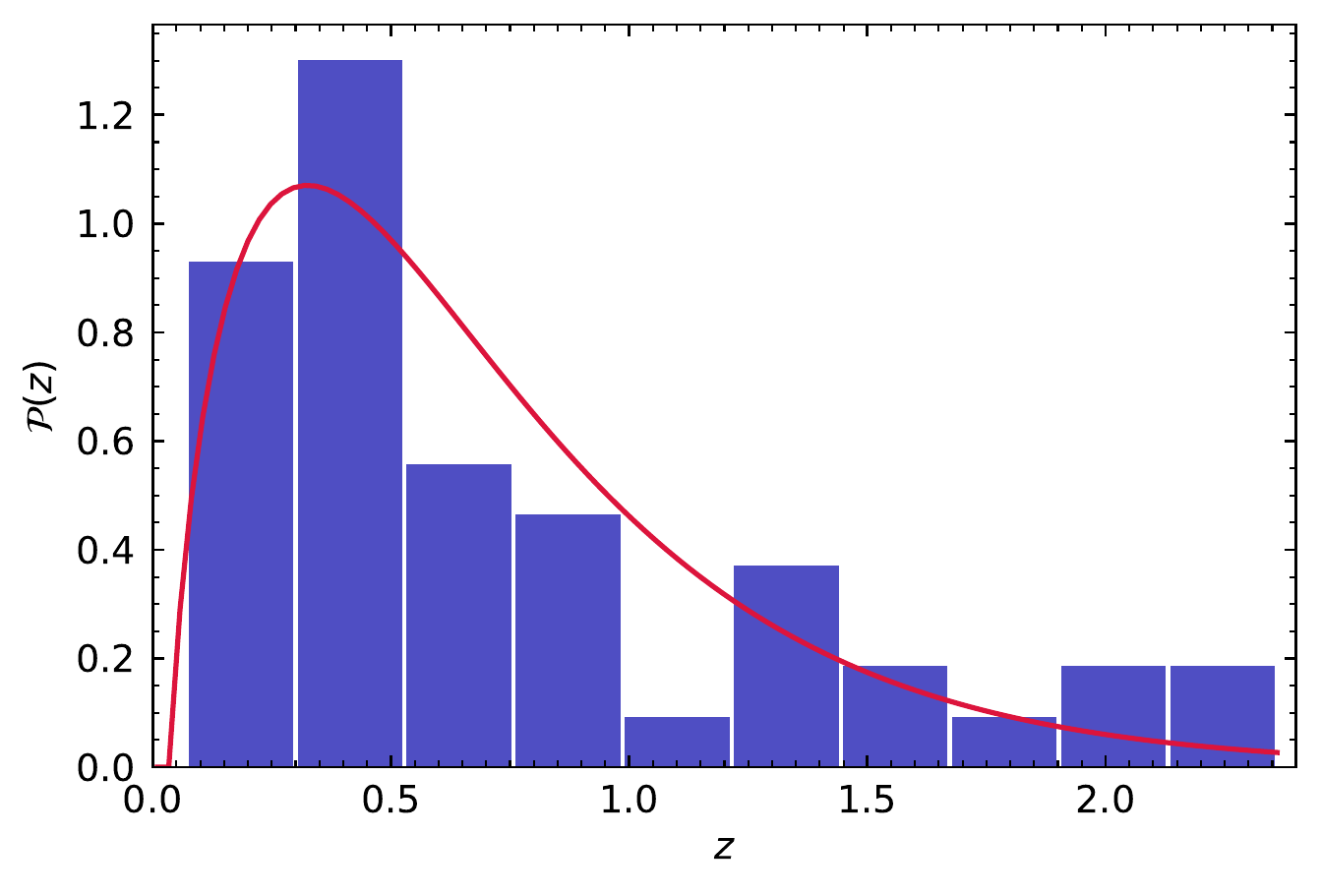}
    \includegraphics[width=0.485\linewidth]{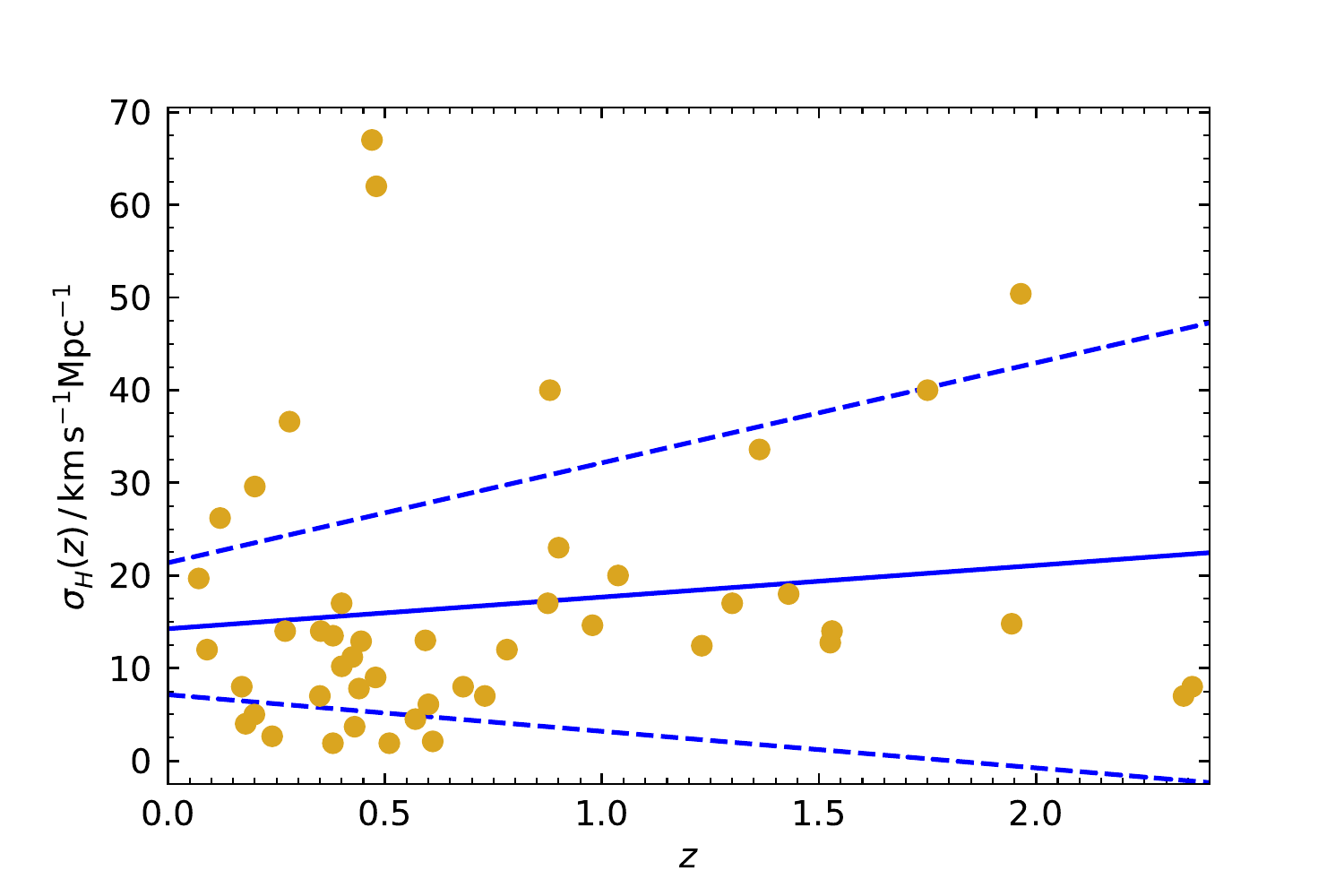}
    \caption{In the left panel we illustrate the distribution of the observational $H(z)$ data points and the assumed redshift distribution function, whereas in the right panel we depict the error of observational $H(z)$ along with their linear regression best-fit.}
    \label{fig:Hz_data}
\end{figure}

The adopted network model was optimised by using forty--seven mock $H(z)$ data points (identical to the number of observational $H(z)$ data points introduced in Sec.~\ref{sec:ANN_Hz_Omz}) which have been simulated in the context of the spatially--flat $\Lambda$CDM model in which 
\begin{equation}
    H(z)=H_0^\mathrm{mock}\sqrt{\Omega_{m,0}^\mathrm{mock}(1+z)^3+1-\Omega_{m,0}^\mathrm{mock}}\,,
\end{equation}
where we assumed that $H_0^{\mathrm{mock}}=70\,\mathrm{km}\,\mathrm{s}^{-1}\mathrm{Mpc}^{-1}$ and $\Omega_{m,0}^\mathrm{mock}=0.3$. We should remark that our final results are independent from the choice of these fiducial cosmological parameter values since we are simply using this model to structure the network rather than actually train it. Furthermore, the redshift distribution of the observational $H(z)$ was assumed to follow a Gamma distribution, specified by
\begin{equation}\label{eq:gamma_dist}
    p(x;\,\alpha,\,\lambda)=\frac{\lambda^\alpha}{\Gamma(\alpha)}x^{\alpha-1}e^{-\lambda x}\,,
\end{equation}
where $\alpha$ and $\lambda$ are free parameters that are fitted with the considered observational data, while the gamma function is given by
\begin{equation}
    \Gamma(\alpha)=\int_0^{\infty} e^{-t}t^{\alpha-1}\,\mathrm{d}t\,.
\end{equation}
This is a well known distribution for these data sets. The distribution of the observational $H(z)$ data points and the fitted redshift distribution function are shown in the left panel of Fig. \ref{fig:Hz_data}.

In order to generate a mock data set of $H(z)$, we also need to take into account the uncertainty of the observational $H(z)$ data points. In the right panel of Fig. \ref{fig:Hz_data}, we illustrate the errors of the considered $H(z)$ data set as a function of redshift. As expected, the uncertainties tend to increase with the redshift. Consequently, for the generation of the $H(z)$ mock data set, we will be assuming a linear model \cite{Ma:2010mr,Velasquez-Toribio:2021ufm} for the error of $H(z)$, in which we are going to fit a first degree polynomial function of redshift. The mean fitting function is found to be $\sigma_H^0(z)=14.25+3.42z$, while the symmetric upper and lower error bands are respectively specified by $\sigma_H^+(z)=21.37+10.79z$ and $\sigma_H^-(z)=7.14-3.95z$. These fitting functions are also depicted in the right panel of Fig. \ref{fig:Hz_data}, in which one could easily observe that the majority of the data points are included in the area enclosed by the $\sigma_H^+(z)$ and $\sigma_H^-(z)$ functions (dashed lines). We again emphasize that the purpose of this exercise is to produce a mock data set that somewhat mimics the observed points so that we can structure our ANN appropriately before proceeding with the training.

We can now randomly generate the error for our $H(z)$ mock data points, in which we assume that the error $\tilde{\sigma}_H^{}(z)$ follows the Gaussian distribution $\mathcal{N}(\sigma_H^0(z),\,\varepsilon_H^{}(z))$, where $\varepsilon_H^{}(z)=(\sigma_H^+(z)-\sigma_H^-(z))/4$, such that $\tilde{\sigma}_H^{}(z)$ falls in the area with a probability of 95\%. Therefore, every simulated Hubble parameter data point $H_\mathrm{sim}^{}(z_i)$ at redshift $z_i$, is computed via $H_\mathrm{sim}^{}(z_i)=H_\mathrm{fid}^{}(z_i)+\Delta H_i$, with the associated uncertainty of $\tilde{\sigma}_H^{}(z_i)$, where $\Delta H_i$ is determined via $\mathcal{N}(0,\,\tilde{\sigma}_H^{}(z_i))$. 

\begin{figure}
    \centering
    \includegraphics[width=0.485\linewidth]{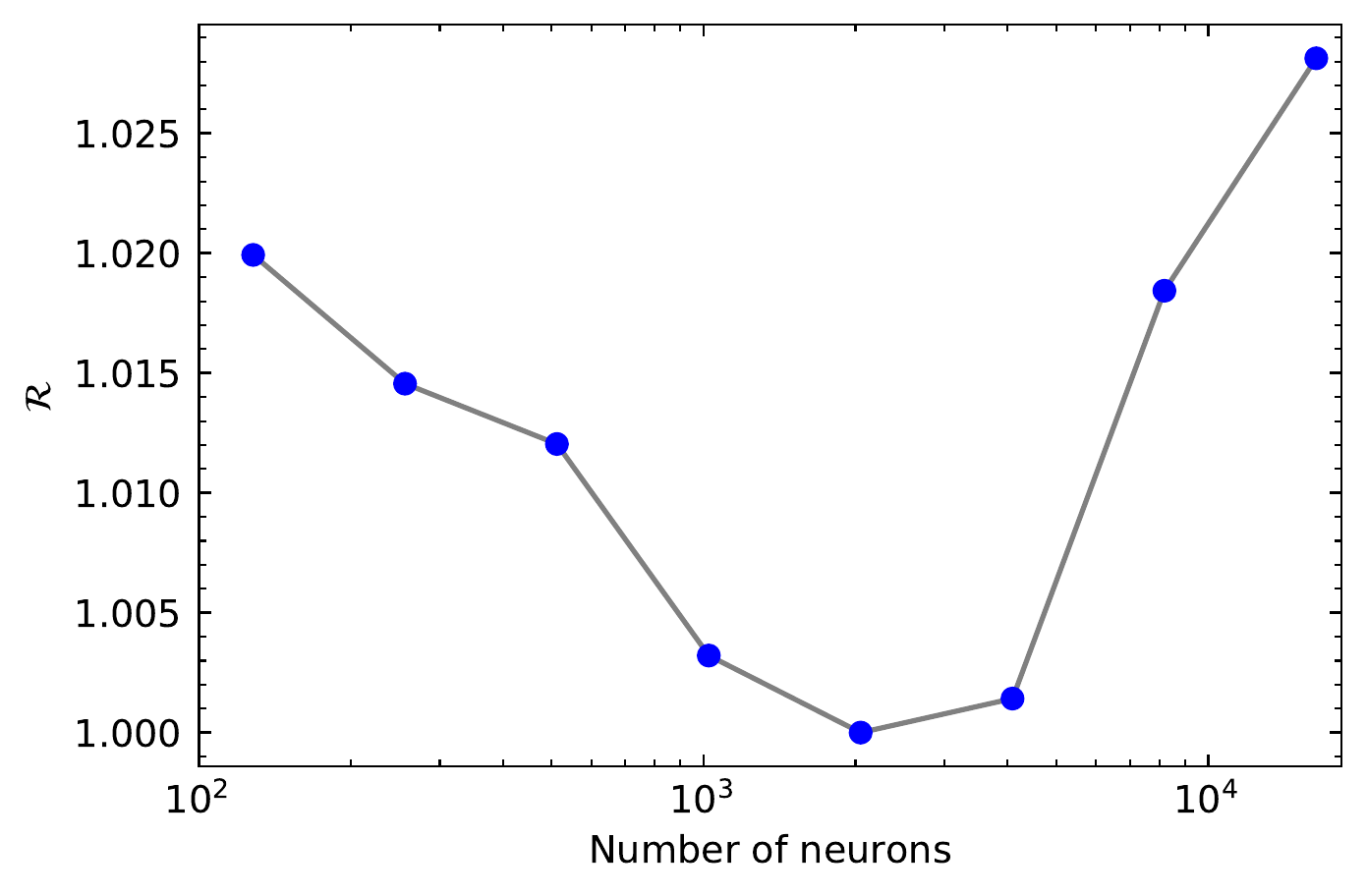}
    \includegraphics[width=0.485\linewidth]{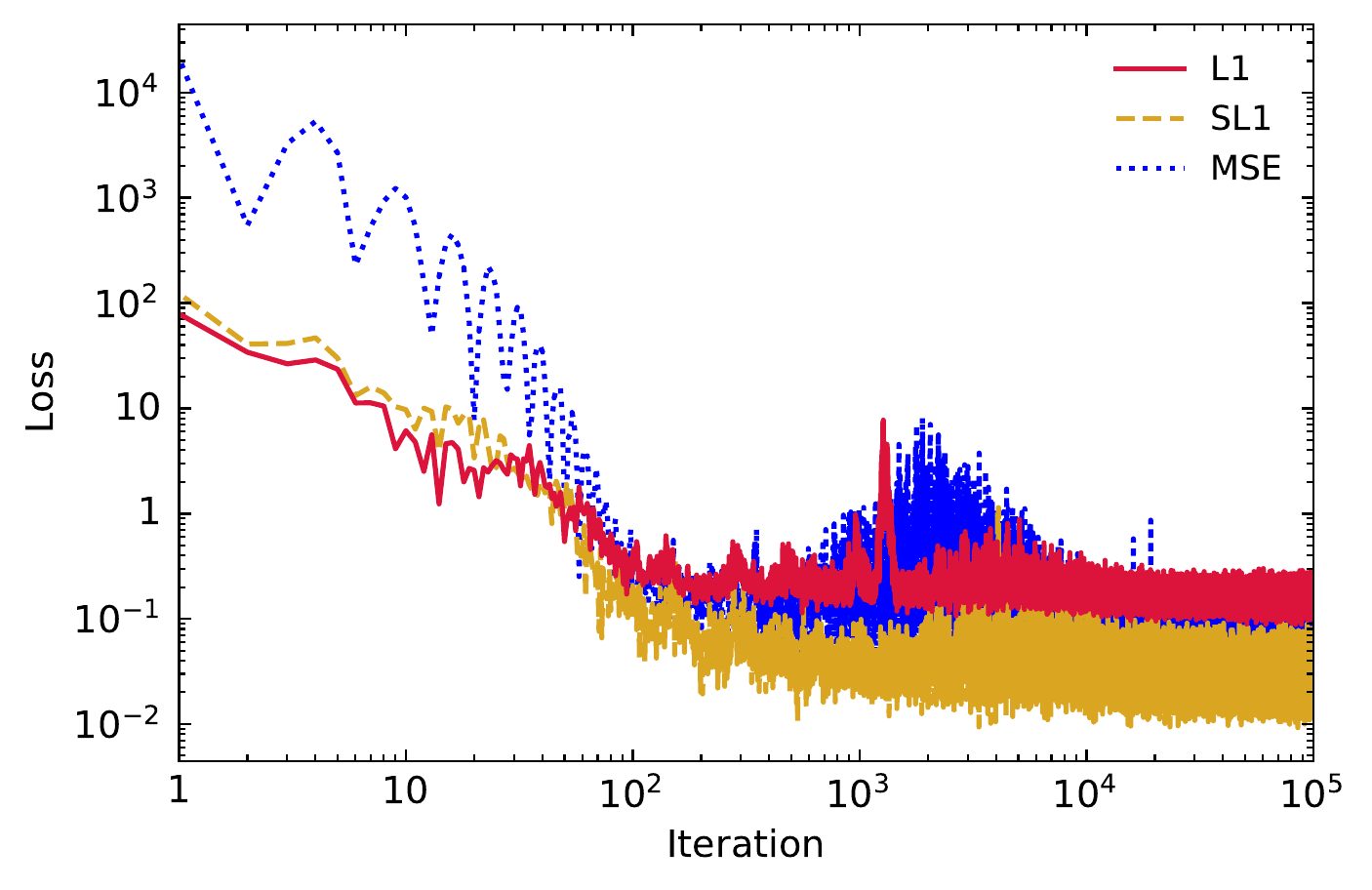}
    \caption{The normalised risk for $H(z)$ ANN models that have one hidden layer and a corresponding number of neurons in their hidden layer is illustrated in the left panel, while the evolution of the L1, SL1 and MSE loss functions is shown in the right panel.}
    \label{fig:Hz_risk_loss}
\end{figure}

Consequently, the data that is adopted to train the network is simulated according to the redshift distribution of the $H(z)$ data points with an assumed spatially--flat $\Lambda$CDM model, and we further consider the same number of mock data points as the number of observational data. As depicted in Fig. \ref{fig:ANN_structure}, the input of the neural network is the redshift, while the output is the corresponding Hubble parameter and its respective error at that redshift. In the training process, the parameters of the neural network will be determined via a learning process using the observational data sets. In our case, the entire $H(z)$ mock data points were used to train the network, and therefore we do not consider a validation and a test set of data like in the standard supervised learning techniques.

For the determination of the optimal network model, we consider finding the optimal number of hidden layers along with the number of neurons in the hidden layers with a selection of loss functions. The initial learning rate is set to 0.01 which will decrease with the number of iterations (which describes the degree to which successive iterations override neuron hyperparameter values), while the training batch size is set to half of the number of the $H(z)$ data points. The network is trained after $10^5$ iterations, such that the loss function no longer decreases after this number of iterations, which is clearly illustrated in the right panel of Fig. \ref{fig:Hz_risk_loss} where the loss function is close to its asymptote.

\begin{figure}
    \centering
    \includegraphics[width=0.485\linewidth]{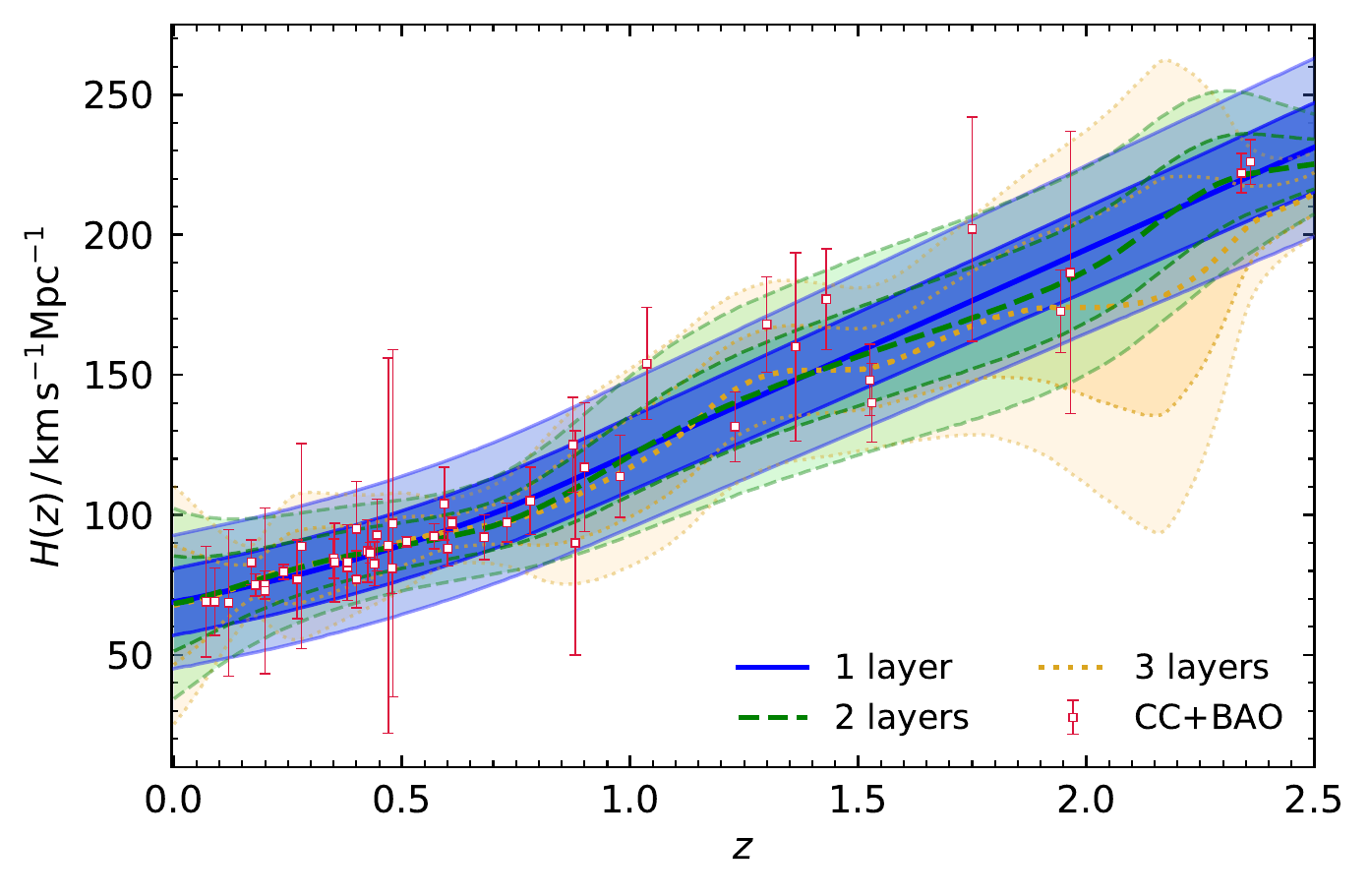}
    \includegraphics[width=0.485\linewidth]{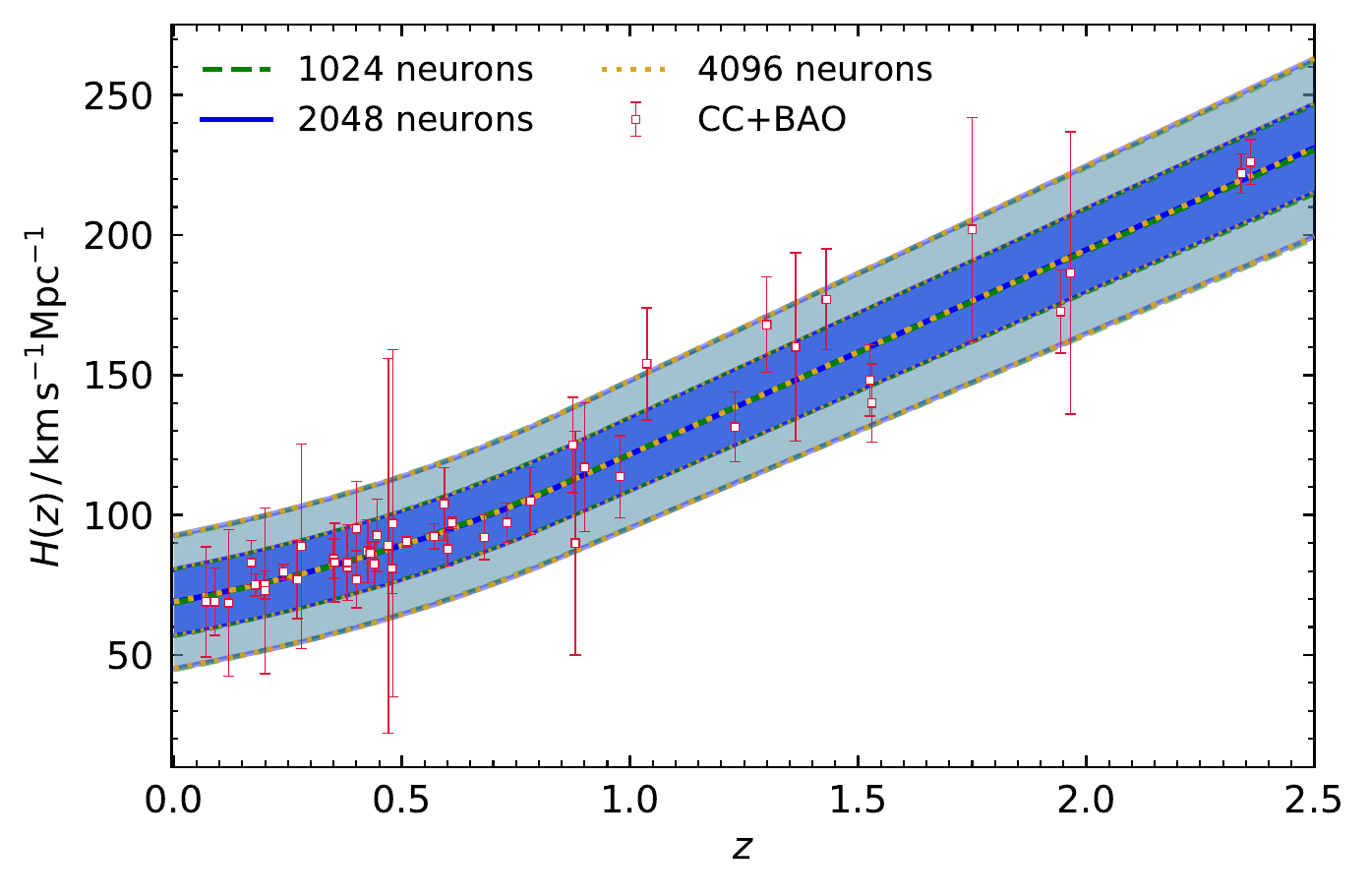}
    \includegraphics[width=0.485\linewidth]{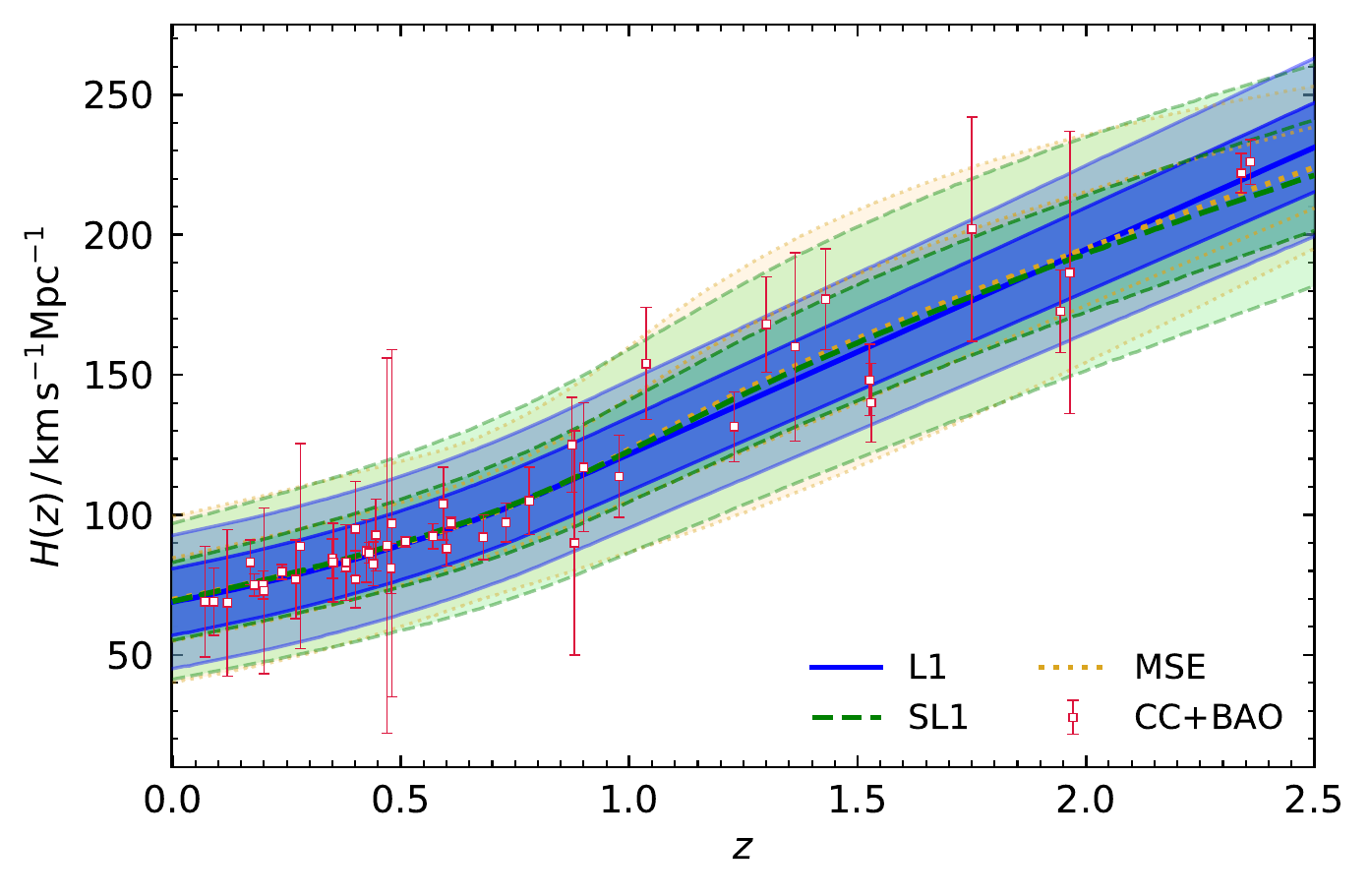}
    \includegraphics[width=0.485\linewidth]{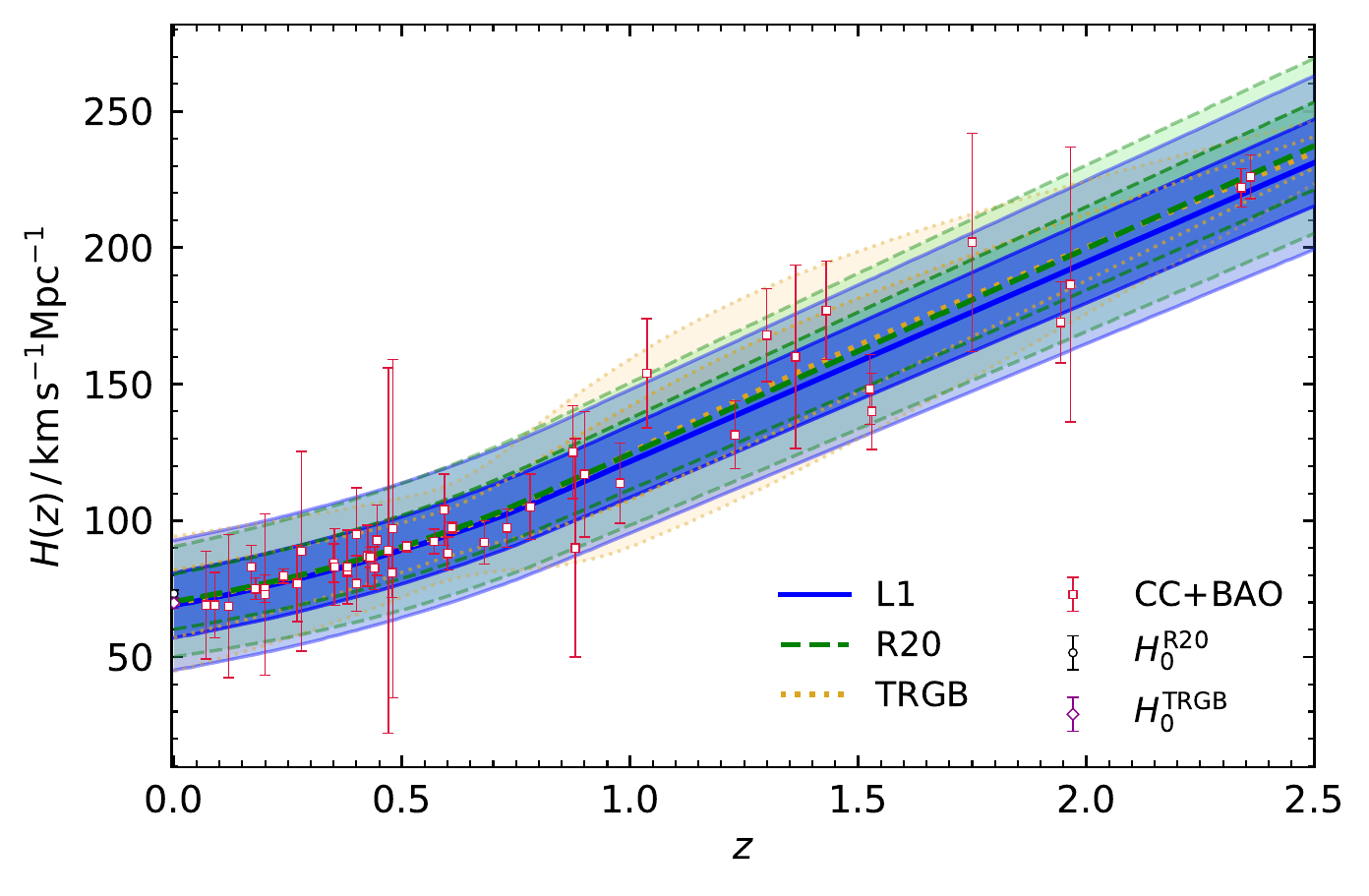}
    \caption{We depict the L1 $H(z)$ ANN reconstructions with different number of layers and neurons in the top-left and top-right panels, respectively. In the bottom-left panel we depict the $H(z)$ ANN reconstructions adapting the L1, SL1 and MSE loss functions, whereas in the bottom-right panel we illustrate the $H(z)$ ANN reconstructions when considering the L1 loss function without an $H_0$ prior (L1), with the R20 prior (R20) and the TRGB prior (TRGB). In all panels, the corresponding $H(z)$ data points are also included.}
    \label{fig:Hz_ANN}
\end{figure}

\begin{figure}
    \centering
    \includegraphics[width=0.785\linewidth]{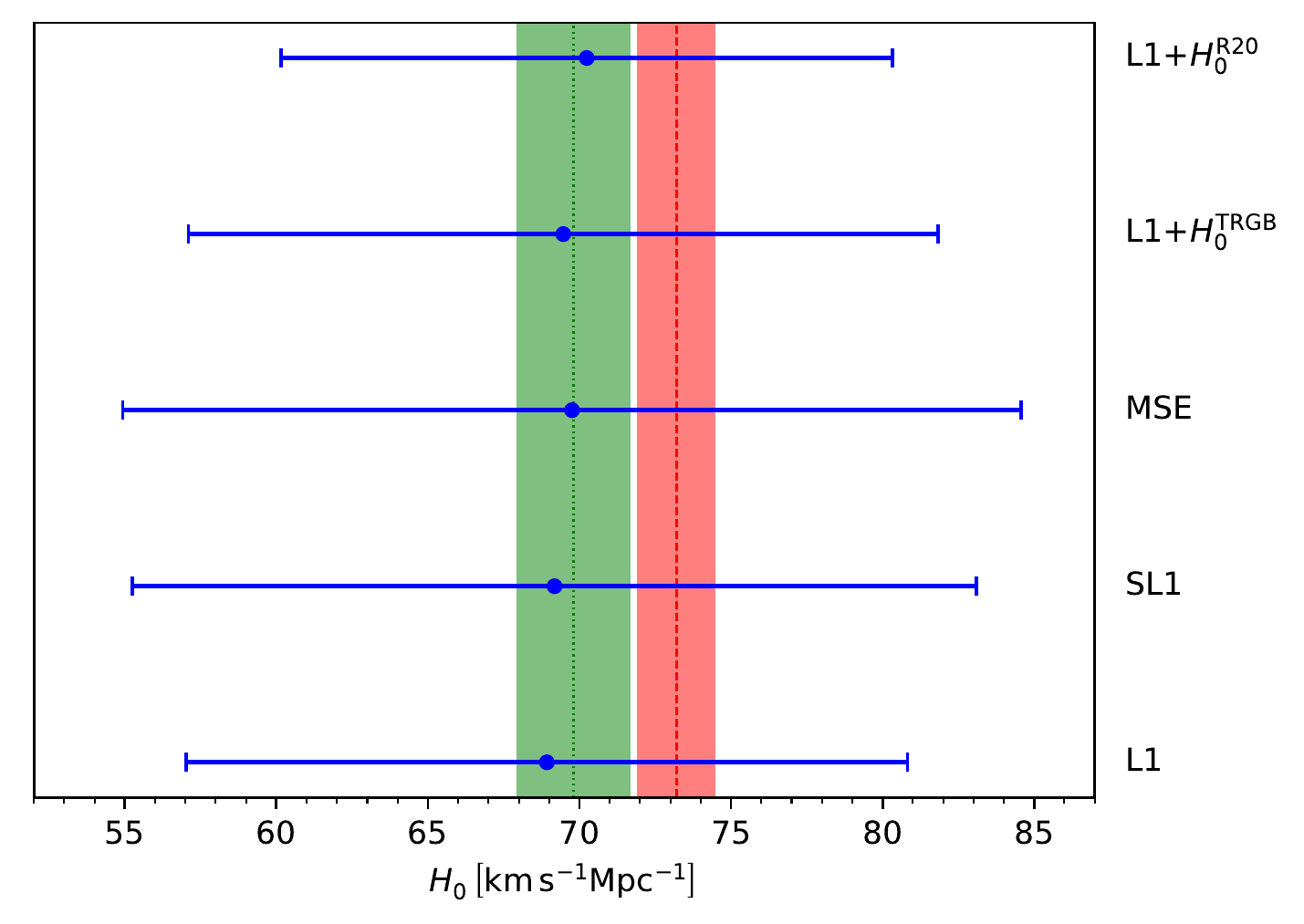}
    \caption{We illustrate the inferred $1\sigma$ constraint on $H_0$ from the $H(z)$ ANN reconstructions as indicated on the vertical axis. The green and red bands illustrate the $1\sigma$ local measurements of $H_0^\mathrm{TRGB}$ and $H_0^\mathrm{R20}$, respectively.}
    \label{fig:H0_ANN}
\end{figure}

For the training procedure, we consider the mock $H(z)$ data set, with the number of hidden layers varying from one to three, and eight network models are trained with $2^n$ number of neurons, where $7\leq n\leq 14$. We therefore train a total of twenty-four network models, from which we will determine the optimal network configuration. This set of trained networks can then be used to select the optimal network structure on which to train the real data. Indeed, to select the optimal number of hidden layers of the network we adopt the risk statistic \cite{Wasserman:2001ng}
\begin{equation}
    \mathrm{risk}=\sum_{i=1}^N\mathrm{bias}_i^2+\sum_{i=1}^N\mathrm{variance}_i^{}=\sum_{i=1}^N\left[H(z_i)-\bar{H}(z_i)\right]^2+\sum_{i=1}^N\sigma^2\left(H(z_i)\right)\,,
\end{equation}
where $N$ is the number of $H(z)$ data points, and $\bar{H}(z)$ denotes the fiducial value of $H(z)$. We then calculate the risk of eight models for each network structure, from which we inferred that the network model with one hidden layer minimises the risk with respect to the two and three hidden layer training networks. We should remark that this was the case for the L1, MSE and SL1 loss functions. The degree of complexity of the ANN should reflect the structure of the physical process which is producing the data. Given that we are using expansion data alone, the lack of complexity in the data seems to infer a simpler one-layer structure to the ANN. This is to be expected since we are only taking $H(z)$ data. The conclusion would naturally be altered if we had coupled this to other cosmological parameters.

We now determine the optimal number of neurons with one hidden layer via the consideration of eight network models with a varying number of neurons. We illustrate the normalised risk values $\mathcal{R}$ for each network model in the left panel of Fig.~\ref{fig:Hz_risk_loss}, in which one could clearly observe that 2048 neurons minimise the risk function. Such a result was also found to be independent from the adopted loss function. Consequently, the network structure with one hidden layer and 2048 neurons was found to be the optimal network structure and will therefore be adopted in our $H(z)$ reconstructions. Furthermore, the L1 loss function shall be adopted in our ANN reconstruction of the observational $H(z)$ data set, since this was characterised by the lowest risk statistic with respect to the MSE and SL1 loss function networks.

\begin{figure}
    \centering
    \includegraphics[width=0.485\linewidth]{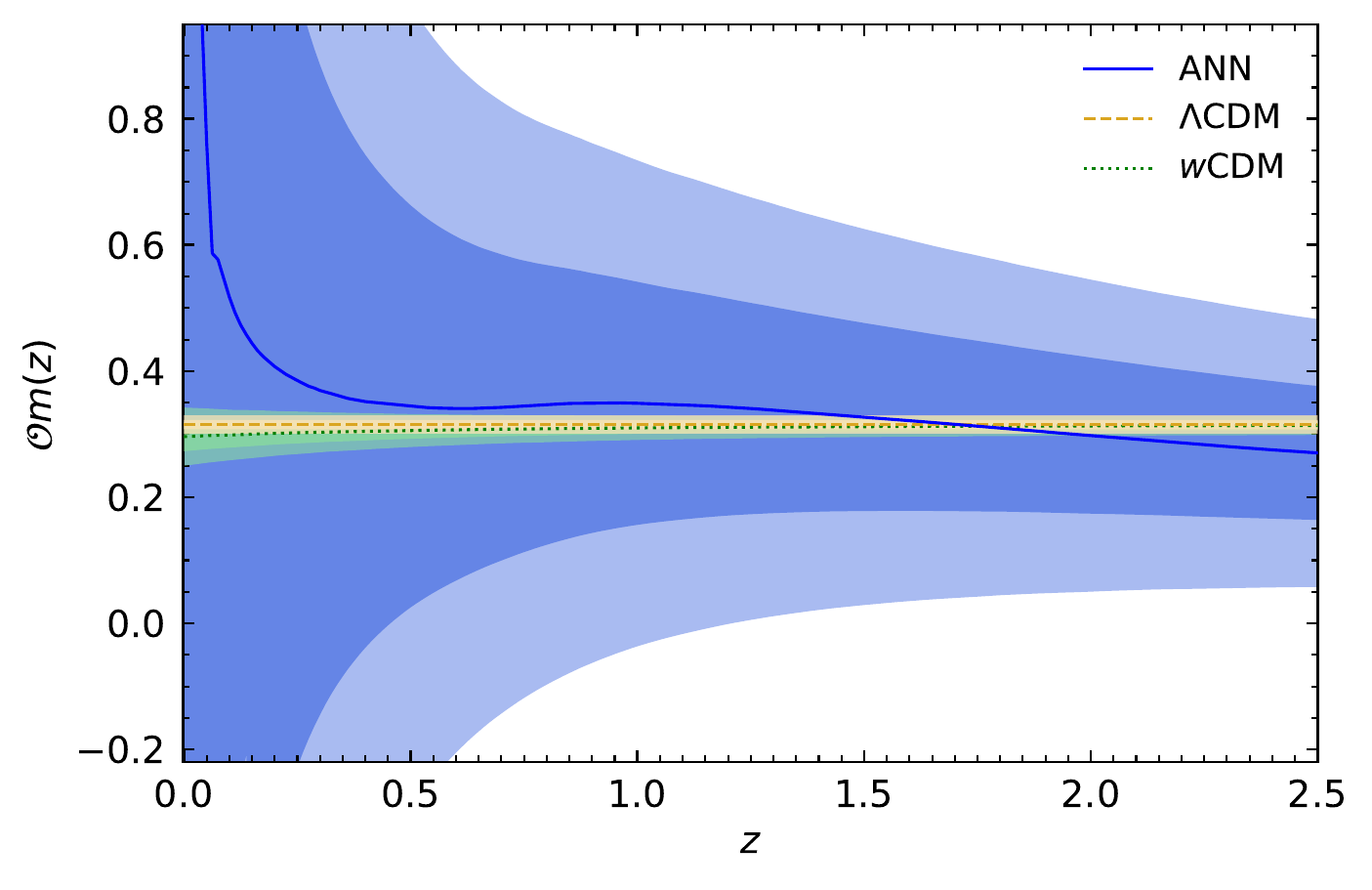}
    \includegraphics[width=0.485\linewidth]{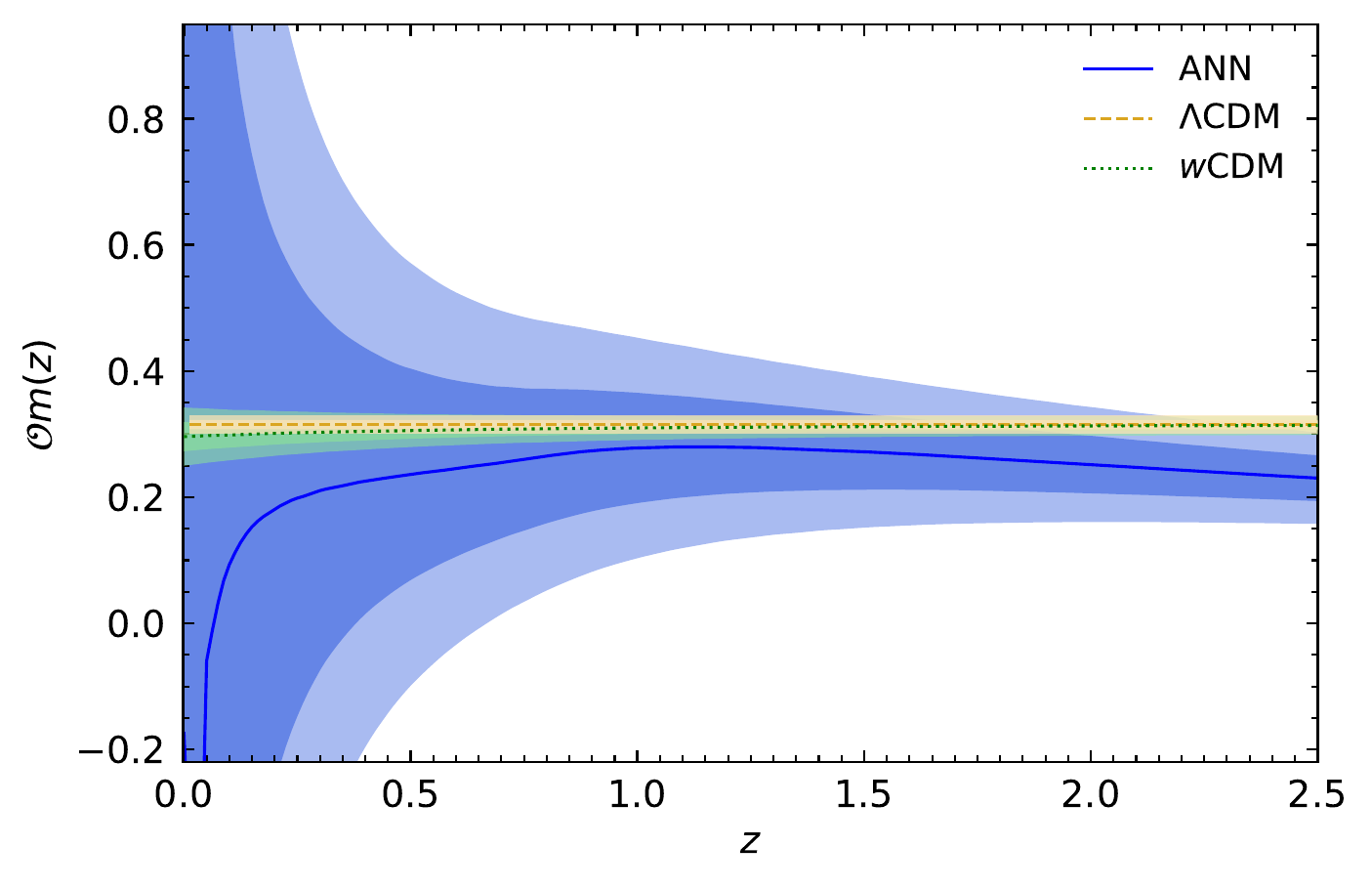}
    \includegraphics[width=0.485\linewidth]{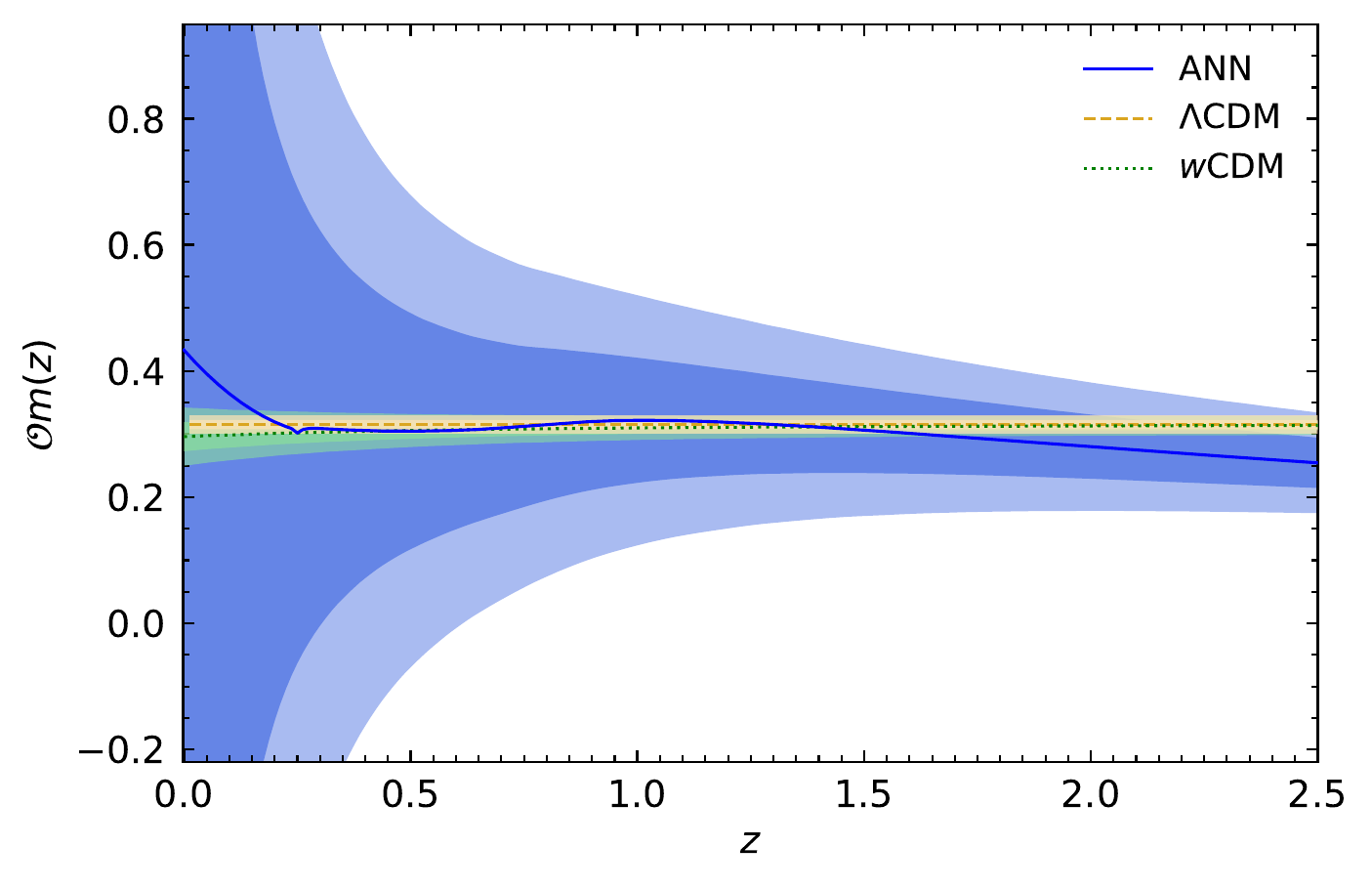}
    \caption{Redshift evolution of the $\mathcal{O}m(z)$ null test without an $H_0$ prior (top--left), with the R20 prior (top--right) and with the TRGB prior (bottom). The \textit{Planck} constraint in the $\Lambda$CDM and $w$CDM frameworks are also illustrated for comparative purposes.}
    \label{fig:Om_z}
\end{figure}

\subsection{\label{sec:ANN_Hz_Omz}ANN \texorpdfstring{$H(z)$}{} reconstructions and \texorpdfstring{$\mathcal{O}m(z)$}{} null test}

Further to the discussion in Sec.~\ref{sec:hubble_training}, an ANN with the L1 loss function having one hidden layer and 2048 neurons will be used for the reconstructions of the $H(z)$ observational data. The considered forty--seven $H(z)$ data points in the range of $0.07< z < 2.36$ were adopted from Refs. \cite{Jimenez:2003iv,Simon:2004tf,Stern:2009ep,Moresco:2012jh,Zhang:2012mp,Moresco:2015cya,Moresco:2016mzx,Ratsimbazafy:2017vga} in the case of cosmic chronometers (CC) \cite{Jimenez:2001gg} data, while Refs. \cite{Zhao:2018gvb,Gaztanaga:2008xz,Blake:2012pj,Samushia:2012iq,Xu:2012fw,BOSS:2014hwf,BOSS:2013igd,BOSS:2016wmc} were consulted in the case of $H(z)$ measurements extracted from the detection of radial baryonic acoustic oscillation (BAO) features under a $\Lambda$CDM prior. Henceforth, we will be referring to this joint CC and BAO $H(z)$ data set by CC$+$BAO. Furthermore, we occasionally make use of two independent local measurements of the Hubble constant, specified by $H_0^\mathrm{R20}=73.2\pm1.3\,\mathrm{km}\,\mathrm{s}^{-1}\mathrm{Mpc}^{-1}$ \cite{Riess:2020fzl} (R20) in the case of Cepheids distance scale and $H_0^\mathrm{TRGB}=69.8\pm1.88\,\mathrm{km}\,\mathrm{s}^{-1}\mathrm{Mpc}^{-1}$ \cite{Freedman:2020dne,Freedman:2019jwv} inferred via the Tip of the Red Giant Branch (TRGB) technique. We use these measurements of $H_0$ to put priors on the Hubble data in the training process.

We illustrate the effect of the number of hidden layers on the ANN reconstruction of the observational $H(z)$ data in the top--left panel of Fig. \ref{fig:Hz_ANN}, in which an increase in the number of hidden layers tend to be characterised by some oscillatory features to accommodate the nearest neighbouring points. As depicted in the top--right panel of Fig. \ref{fig:Hz_ANN}, different number of neurons led to a minute difference in the reconstructions, while different loss functions are characterised by different reconstructions of $H(z)$, as shown in the bottom--left panel of Fig. \ref{fig:Hz_ANN}. The inferred Hubble constant constraints with the MSE, L1, and SL1 loss functions were found to be $H_0=69.76\pm14.82\,\mathrm{km}\,\mathrm{s}^{-1}\mathrm{Mpc}^{-1}$, $H_0=68.93\pm11.90\,\mathrm{km}\,\mathrm{s}^{-1}\mathrm{Mpc}^{-1}$, and $H_0=69.18\pm13.92\,\mathrm{km}\,\mathrm{s}^{-1}\mathrm{Mpc}^{-1}$, respectively. 
Indeed, the variation in the ANN determination of $H_0$ was found to be independent from the adopted loss function, as illustrated in Fig. \ref{fig:H0_ANN}. These characteristics differ from the GP technique in which the extrapolated Hubble parameter to redshift zero is known to be dependent on the chosen kernel function \cite{Briffa:2020qli} and also on the range of hyperparameter values \cite{Sun:2021pbu}.

We also analyse the effect of two Hubble constant prior values on our ANN determination of $H_0$ and the redshift evolution of the ANN reconstructed $H(z)$ function. The ANN reconstructions with the inclusion of the R20 and the TRGB priors are illustrated in the bottom--right panel of Fig. \ref{fig:Hz_ANN}, in which we also compare with the $H(z)$ ANN reconstruction without an $H_0$ prior. The derived Hubble constant constraints were found to be $H_0=70.24\pm10.08\,\mathrm{km}\,\mathrm{s}^{-1}\mathrm{Mpc}^{-1}$ with the R20 prior, and $H_0=69.47\pm12.37\,\mathrm{km}\,\mathrm{s}^{-1}\mathrm{Mpc}^{-1}$ with the TRGB prior. When compared with the L1 reconstructed value of $H_0=68.93\pm11.90\,\mathrm{km}\,\mathrm{s}^{-1}\mathrm{Mpc}^{-1}$, we could observe that the mean value of the Hubble constant was affected by each prior, although the constraints are in an excellent agreement with one another, as illustrated in Fig. \ref{fig:H0_ANN}. Moreover, the reconstructed $H(z)$ functions are very similar to each other as well. Hence, the ANN reconstructions of $H(z)$ are nearly independent on an $H_0$ prior. This outcome has also been reported in Ref. \cite{Wang:2019vxv}, which distinguishes the ANN method of reconstruction from other machine learning techniques, such as GP in which an $H_0$ prior affects the reconstructed functions \cite{Briffa:2020qli}.

We now consider a cosmic expansion diagnostic test of the $\Lambda$CDM model which is formulated in terms of the Hubble parameter function, that simplifies to a constant if the Universe is exactly described by the $\Lambda$CDM model. Deviations from the concordance model of cosmology via the reconstruction of cosmographic functions \cite{Velasquez-Toribio:2021ufm,Reyes:2021owe} and null tests \cite{Zunckel:2008ti,Sahni:2008xx,Shafieloo:2009hi,Clarkson:2007pz,Qi:2016wwb,Qi:2018pej,Bengaly:2020neu} have been exhaustively explored in the literature. The considered cosmic expansion consistency relation is given by \cite{Sahni:2008xx,Shafieloo:2009hi}
\begin{equation}\label{eq:Om_z}
    \mathcal{O}m(z)=\frac{E^2(z)-1}{(1+z)^3-1}\,,
\end{equation}
where $E^2(z)=H^2(z)/H_0^2$. In the specific case of the $\Lambda$CDM model, this diagnostic function reduces to the matter density parameter $\Omega_{m}^0$ independent of the redshift. Hence, any non--constant evolution of $\mathcal{O}m(z)$ could be an indication of modified gravity or any other dark energy modification of the concordance model of cosmology. For instance, for
the most elementary phenomenology of dark energy parametrised via a constant dark energy equation of state parameter $w$ within the framework of the $w$CDM model, a positive slope of the $\mathcal{O}m(z)$ diagnostic function is a characteristic of a phantom equation of state $(w<-1)$, while a negative slope is related to the quintessence dark energy model specified by $w>-1$.

\begin{figure}
    \centering
    \includegraphics[height=0.329\linewidth]{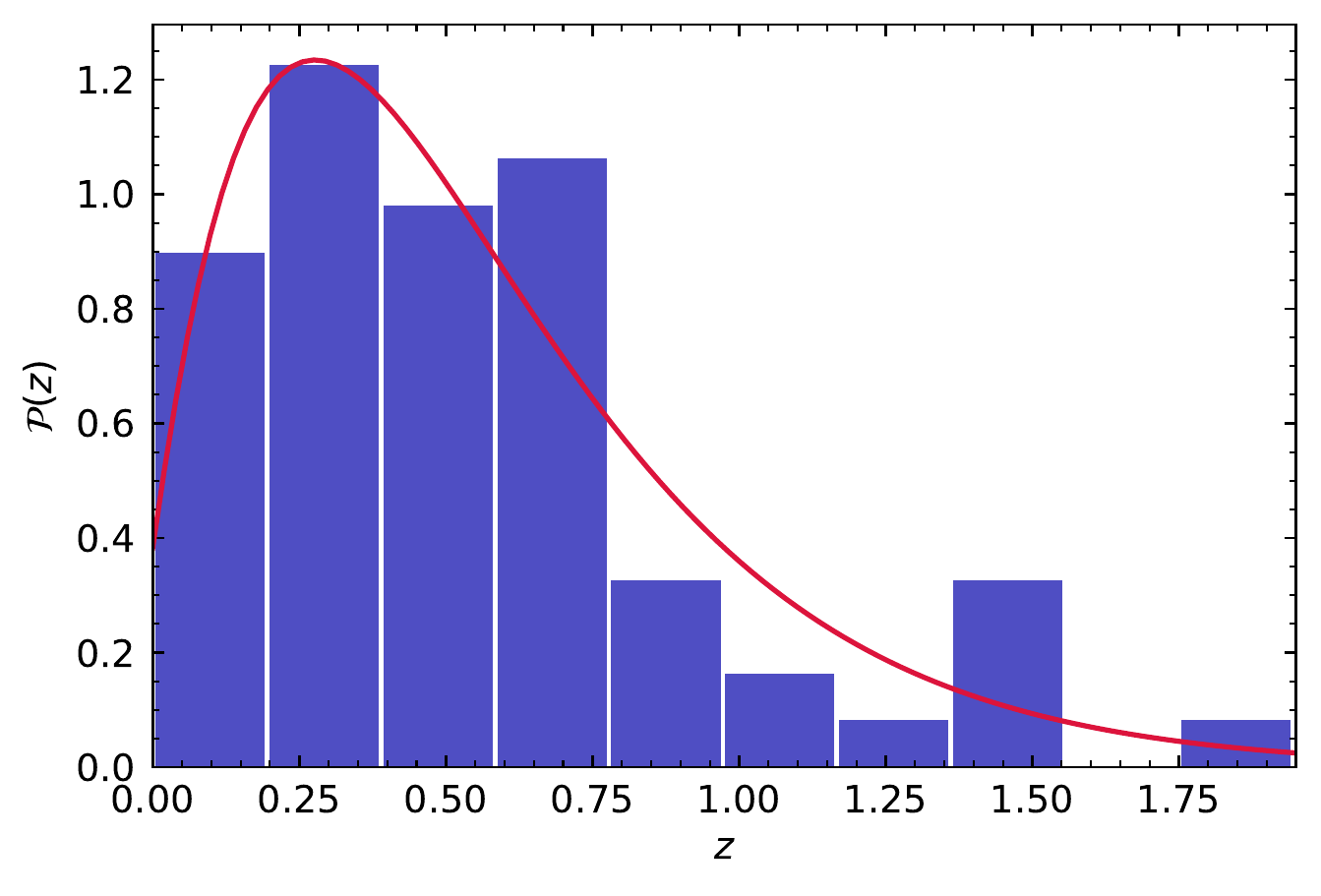}
    \includegraphics[height=0.329\linewidth]{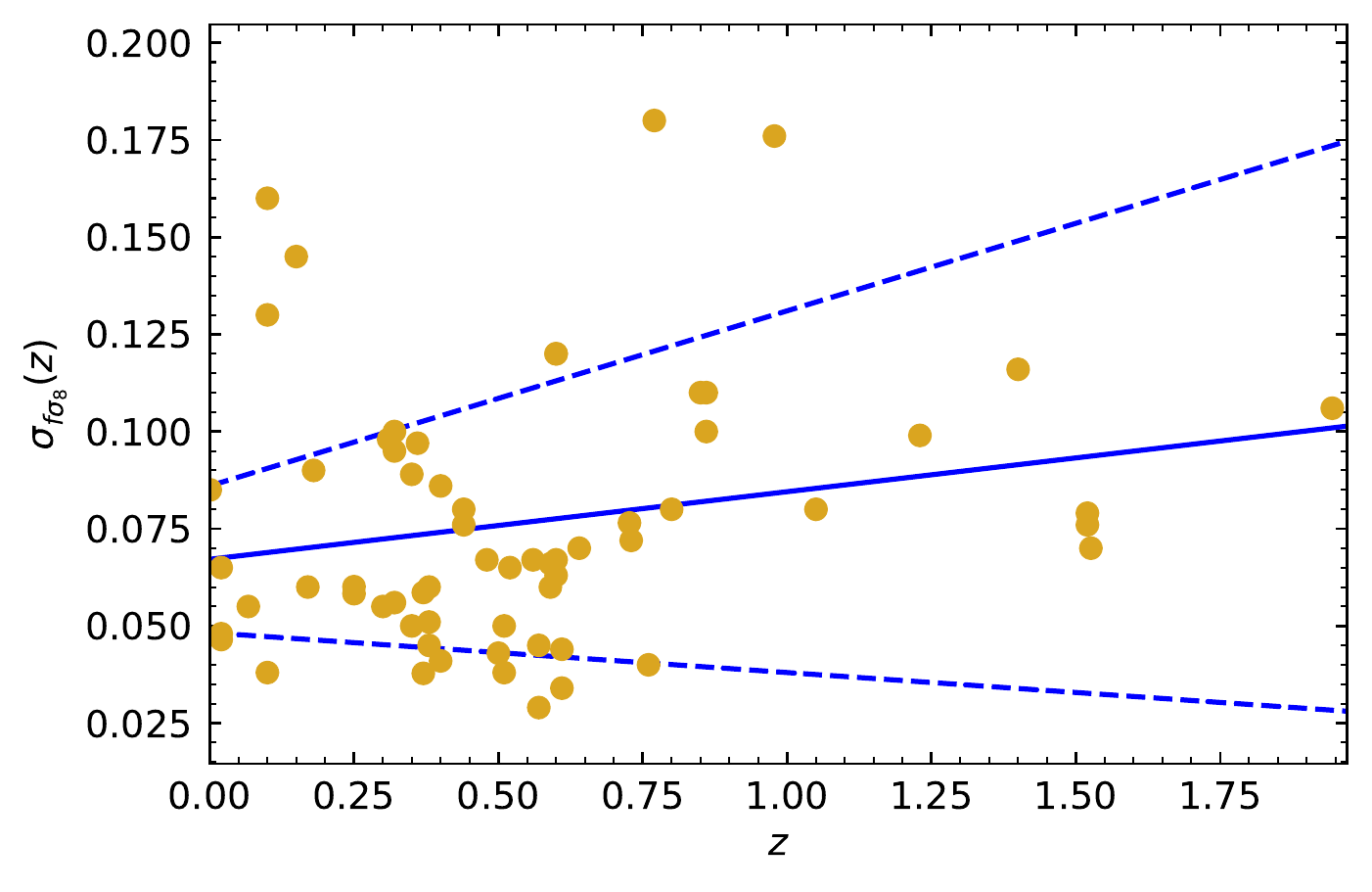}
    \caption{In the left panel we illustrate the distribution of the observational $f\sigma_8(z)$ data points and the assumed redshift distribution function, whereas in the right panel we depict the error of observational $f\sigma_8(z)$ along with their linear regression best-fit.}
    \label{fig:fs8z_data}
\end{figure}

We have used the inferred ANN $H(z)$ reconstructions to determine nonparametric model--independent reconstructions of the $\mathcal{O}m(z)$ function by using Eq. (\ref{eq:Om_z}). In the panels of Fig. \ref{fig:Om_z} we show the results of the ANN reconstructions of the considered cosmic expansion diagnostic function. In the top--left panel we have not assumed an $H_0$ prior, while in the other panels of Fig. \ref{fig:Om_z} we adopted the R20 and TRGB priors. One could easily observe that when we incorporate an $H_0$ prior, we get an improvement in the derived evolution of the ANN reconstruction, particularly at higher redshifts. This occurs due to the extra point at low redshift which has small uncertainties, which then propagate to the high redshift region. At low redshifts $(z\lesssim0.3)$, the ANN reconstruction is not well constrained, although at higher redshifts the profile of the ANN $\mathcal{O}m(z)$ reconstruction tends to be characterised by a negative slope, although an improvement in the data is required in order to distinguish between a preference to either phantom or non--phantom dark energy models (see, for instance, Refs. \cite{Qi:2016wwb,Qi:2018pej,Bengaly:2020neu} for similar conclusions). In Fig. \ref{fig:Om_z} we also illustrate the redshift evolution of the cosmic expansion null test within the $\Lambda$CDM and $w$CDM models when adopting the current \textit{Planck} constraints \cite{Aghanim:2018eyx}. All ANN reconstructions are in good agreement with the $\Lambda$CDM and $w$CDM models, although at $z\gtrsim2$ the ANN reconstructions deviate slightly from these models when adopting the TRGB or R20 $H_0$ priors.

\section{\label{sec:fs8}Cosmic growth data}
We now focus on the cosmological growth evolution via observational $f\sigma_8^{}(z)$ data. We would be following the outlined ANN training procedure of Sec.~\ref{sec:hubble_training}, in which we will be generating $f\sigma_8^{}(z)$ mock data points in Sec.~\ref{sec:fs8_training}, that we will then use for the ANN reconstruction of observational $f\sigma_8^{}(z)$ data. We will further consider a diagnostic test of the concordance model of cosmology at the end of Sec.~\ref{sec:ANN_fs8z_Omz}.

\subsection{\label{sec:fs8_training}Simulation and training of \texorpdfstring{$f\sigma_8^{}(z)$}{} data}

\begin{figure}
    \centering
    \includegraphics[width=0.485\linewidth]{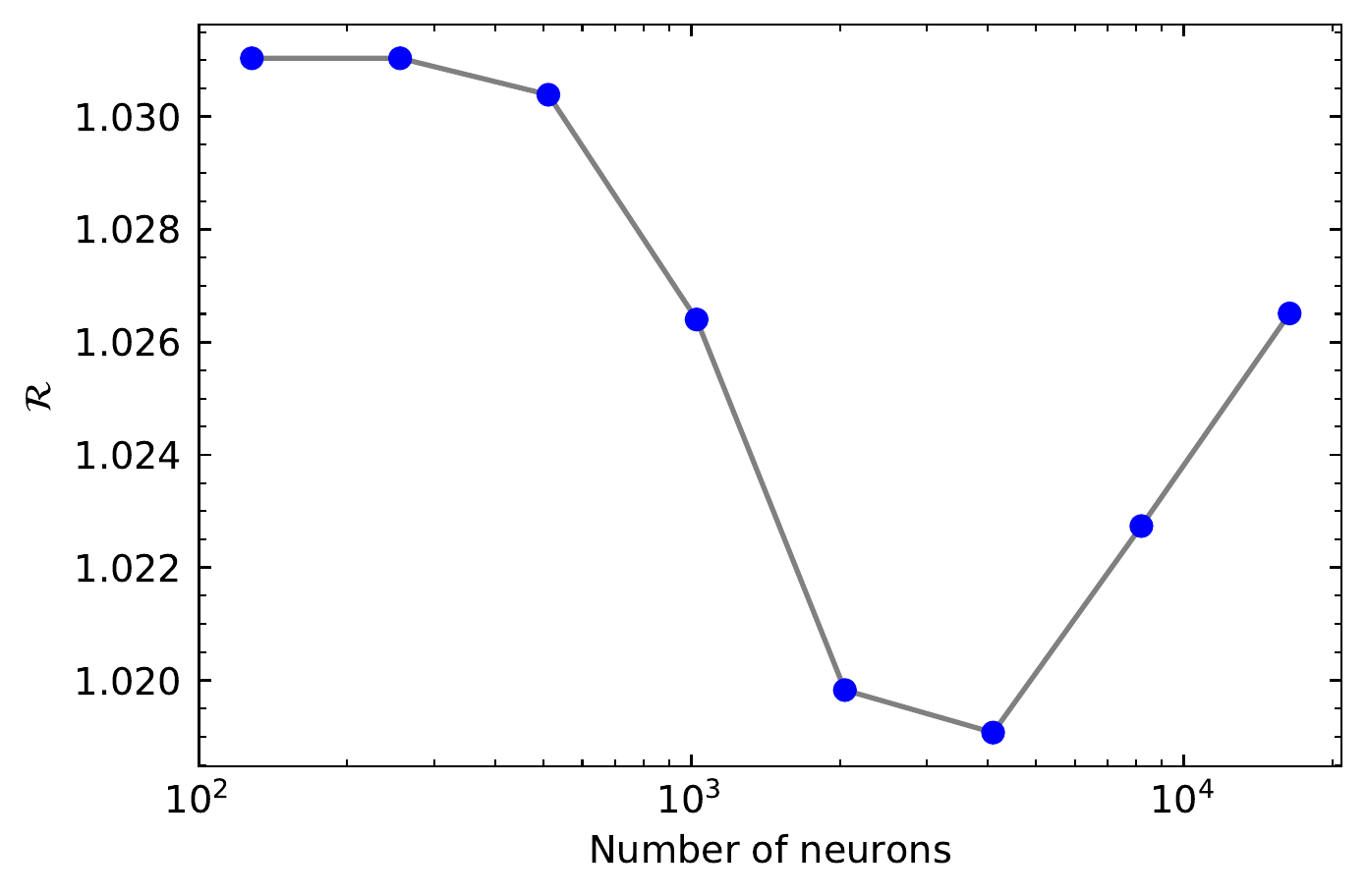}
    \includegraphics[width=0.485\linewidth]{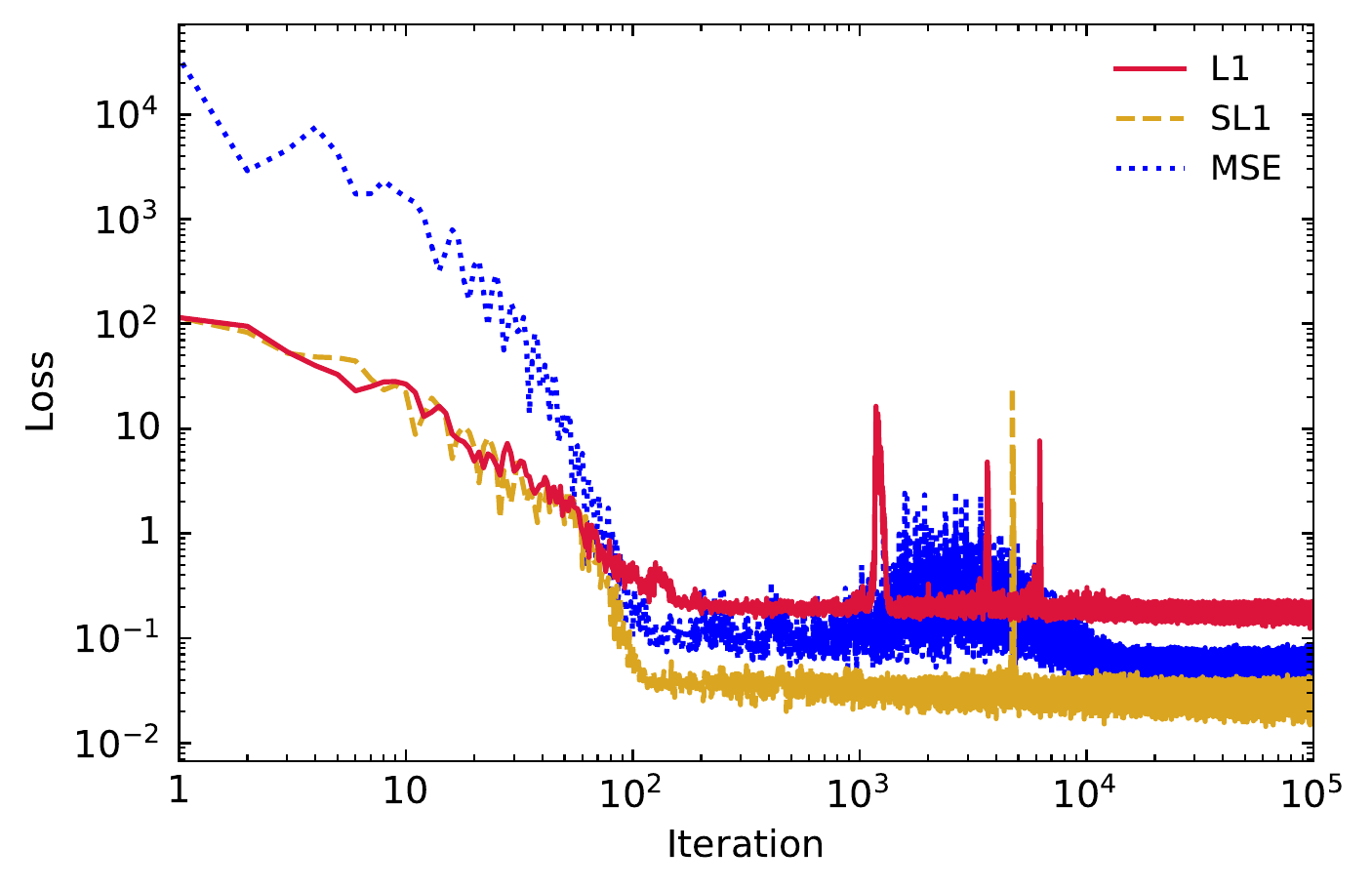}
    \caption{The normalised risk for $f\sigma_8(z)$ ANN models that have one hidden layer and a corresponding number of neurons in their hidden layer is illustrated in the left panel, while the evolution of the L1, SL1 and MSE loss functions is shown in the right panel.}
    \label{fig:fs8_risk_loss}
\end{figure}

The adopted network model was optimised by using sixty--three mock $f\sigma_8(z)$ data points, which is identical to the number of observational data points as summarised in Ref. \cite{Kazantzidis:2018rnb} that were originally reported in Refs. \cite{Blake:2012pj,Beutler:2012px,delaTorre:2013rpa,Chuang:2012qt,WMAP:2010qai,Blake:2013nif,Sanchez:2013tga,BOSS:2013rlg,Howlett:2014opa,Feix:2015dla,Okumura:2015lvp,WMAP:2012nax,BOSS:2016psr,Gil-Marin:2016wya,Huterer:2016uyq,Pezzotta:2016gbo,Howlett:2017asq,Mohammad:2017lzz,Shi:2017qpr,Zhao:2018gvb,Gil-Marin:2018cgo,BOSS:2016wmc,Wang:2017wia,Hou:2018yny,Feix:2016qhh}. For the generated $f\sigma_8(z)$ data points, the spatially--flat $\Lambda$CDM model will be adopted, in which the evolution of the linear growth rate $f(a)$, with respect to the scale factor $a=(1+z)^{-1}$, is governed by the equation
\begin{equation}
    \frac{\mathrm{d}f(a)}{\mathrm{d}\ln a}+f^2(a)+\left(2+\frac{1}{2}\frac{\mathrm{d}\ln H^2(a)}{\mathrm{d}\ln a}\right)f(a)-\frac{3}{2}\Omega_m(a)=0\,,
\end{equation}
where $\Omega_m(a)=\Omega_{m,0}a^{-3}H_0^2/H^2(a)$, and  $f(a)\equiv\mathrm{d}\ln\delta_m(a)/\mathrm{d}\ln a$ is the logarithmic derivative of the growth of matter perturbations $\delta_m\equiv\delta\rho_m/\rho_m$. We should remark that this equation is applicable on sub--horizon scales and in the linear regime. We will now consider the evolution of the matter density contrast $\delta_m(a)$, which is governed by 
\begin{equation}
    \delta_m^{\prime\prime}(a)+\left(\frac{3}{a}+\frac{H^\prime(a)}{H(a)}\right)\delta_m^\prime(a)-\frac{3}{2}\frac{\Omega_m(a)}{a^2}\delta_m(a)=0\,,
\end{equation}
where the prime denotes a derivative with respect to the scale factor. Consequently, the evolution of $\delta_m(a)$ could be specified in terms of the Gaussian hyper--geometric function ${}_{2}F_{1}(a,\,b;\,c;\,d)$ \cite{abramowitz1988handbook}, given by
\begin{equation}
    \delta_m(a)=a\,{}_{2}F_{1}\left[\frac{1}{3},1;\frac{11}{6};a^3\left(1-\frac{1}{\Omega_{m,0}}\right)\right]\,.
\end{equation}
Since the RSD data is expressed in terms of $f\sigma_8(z)=f(z)\sigma_8(z)$, we shall now consider this quantity as follows
\begin{equation}
    f\sigma_8^{}(a)=a\frac{\delta_m^\prime(a)}{\delta_m(a_0)}\sigma_{8,0}^{}\,,
\end{equation}
such that the present day linear theory amplitude of matter fluctuations averaged in spheres of radius $8\,h^{-1}\mathrm{Mpc}$ is specified by \cite{Arjona:2020kco}
\begin{equation}
    \sigma_{8,0}^{}=\int_0^1\frac{f\sigma_8(x)}{x}\mathrm{d}x\,.
\end{equation}
For our mock data set, we considered $\sigma_{8,0}^{\mathrm{mock}}=0.8$ and $\Omega_{m,0}^\mathrm{mock}=0.3$, which are in agreement with the current observational constraints \cite{Aghanim:2018eyx}. It should be emphasised that our final results were found to be independent from the choice of these fiducial cosmological parameter values. Similar to the case of cosmic expansion data, we shall be assuming a Gamma distribution, as specified in Eq. (\ref{eq:gamma_dist}), for the redshift distribution of the observational $f\sigma_8(z)$ data points, which we also illustrate in the left panel of Fig. \ref{fig:fs8z_data}.

\begin{figure}
    \centering
    \includegraphics[width=0.485\linewidth]{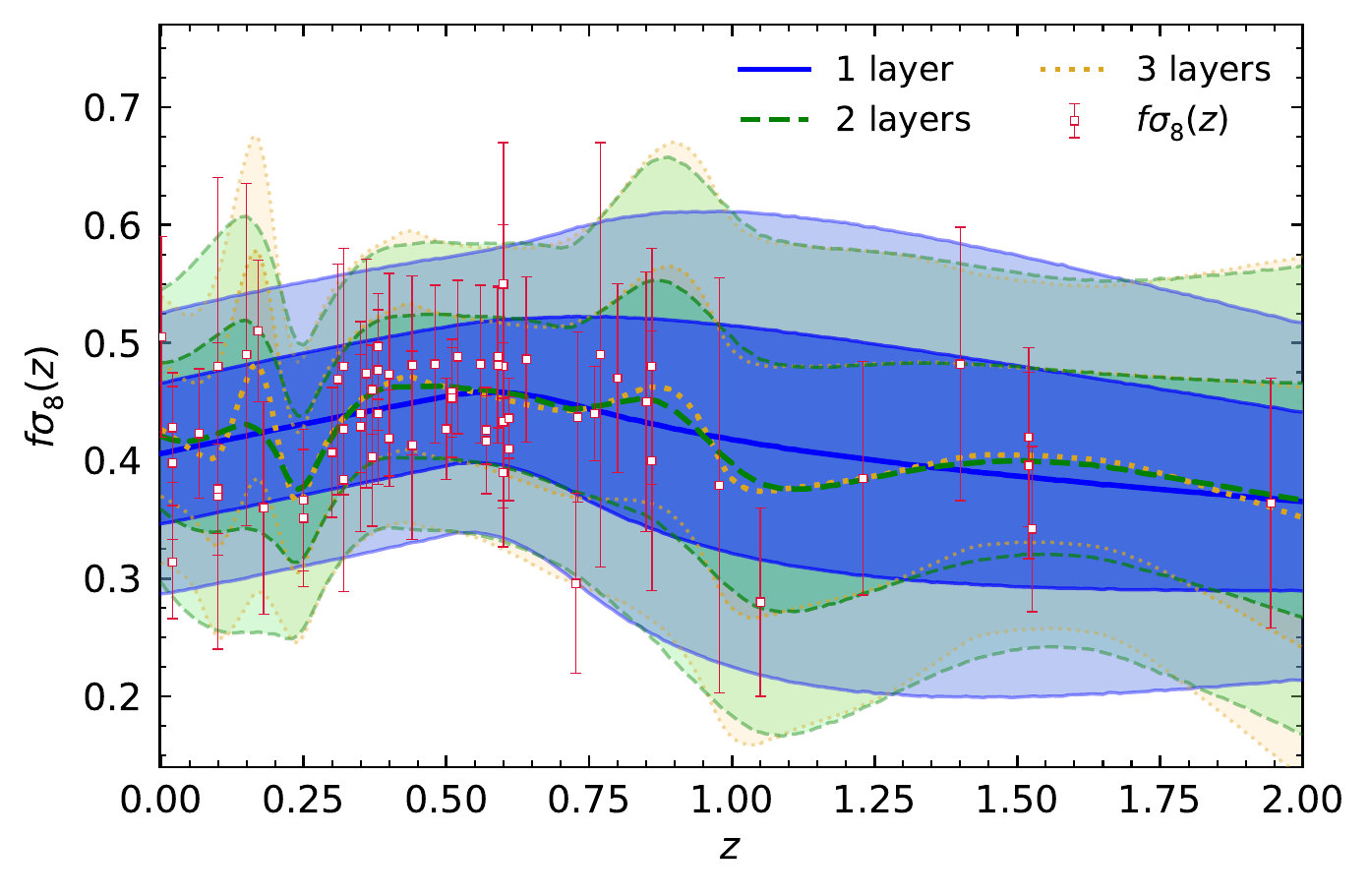}
    \includegraphics[width=0.485\linewidth]{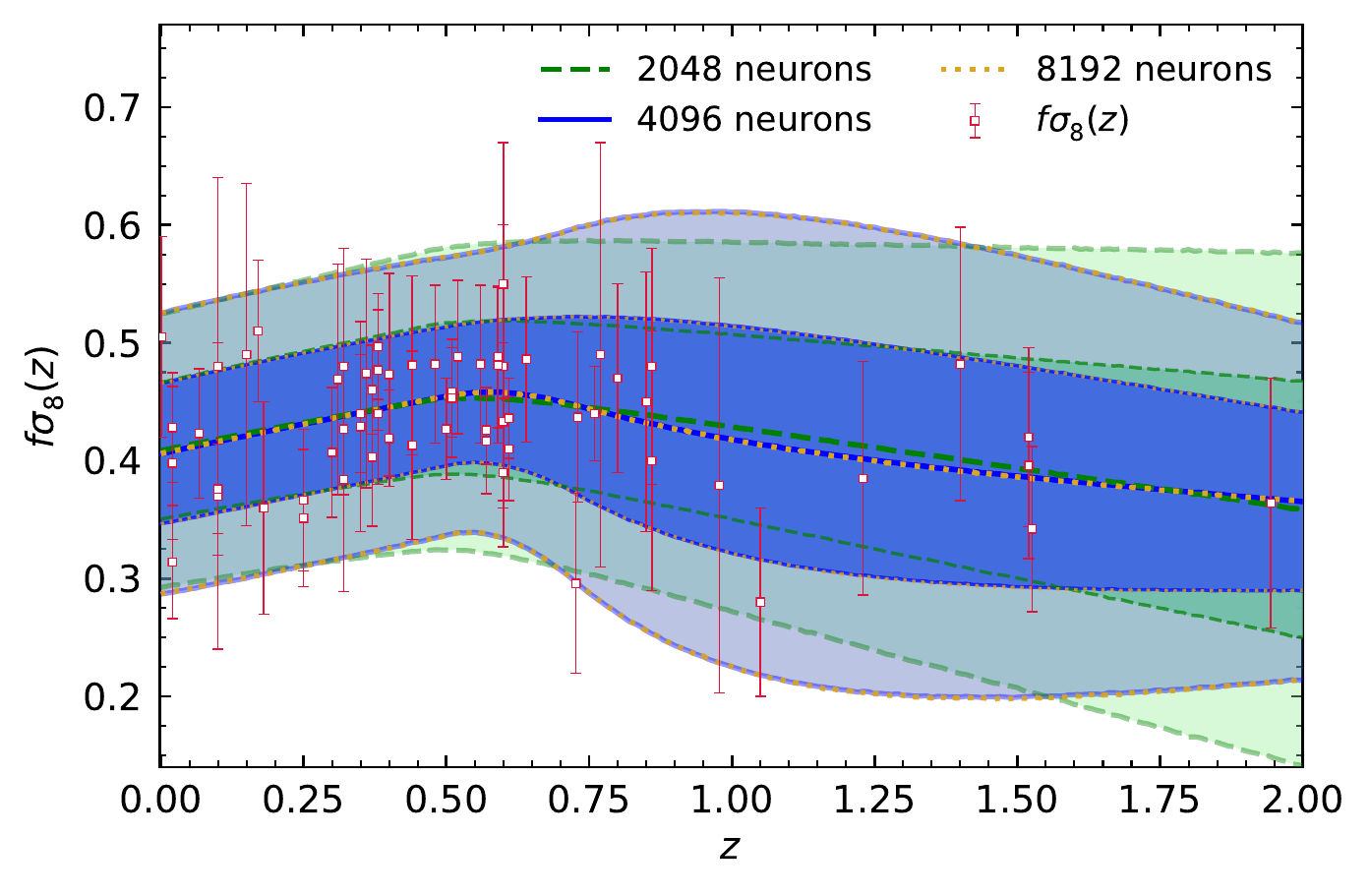}
    \includegraphics[width=0.4865\linewidth]{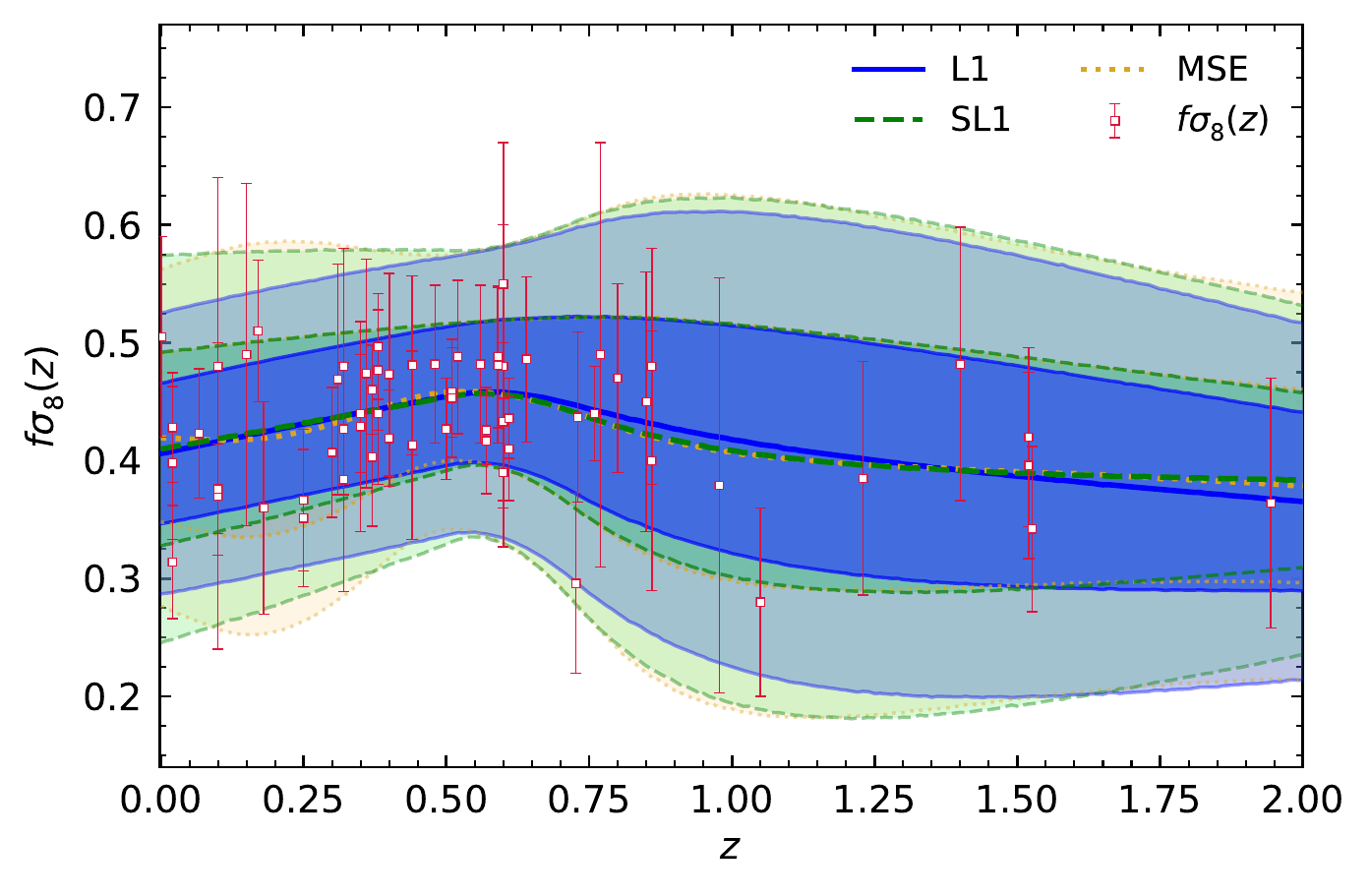}
    \includegraphics[height=0.3555\linewidth]{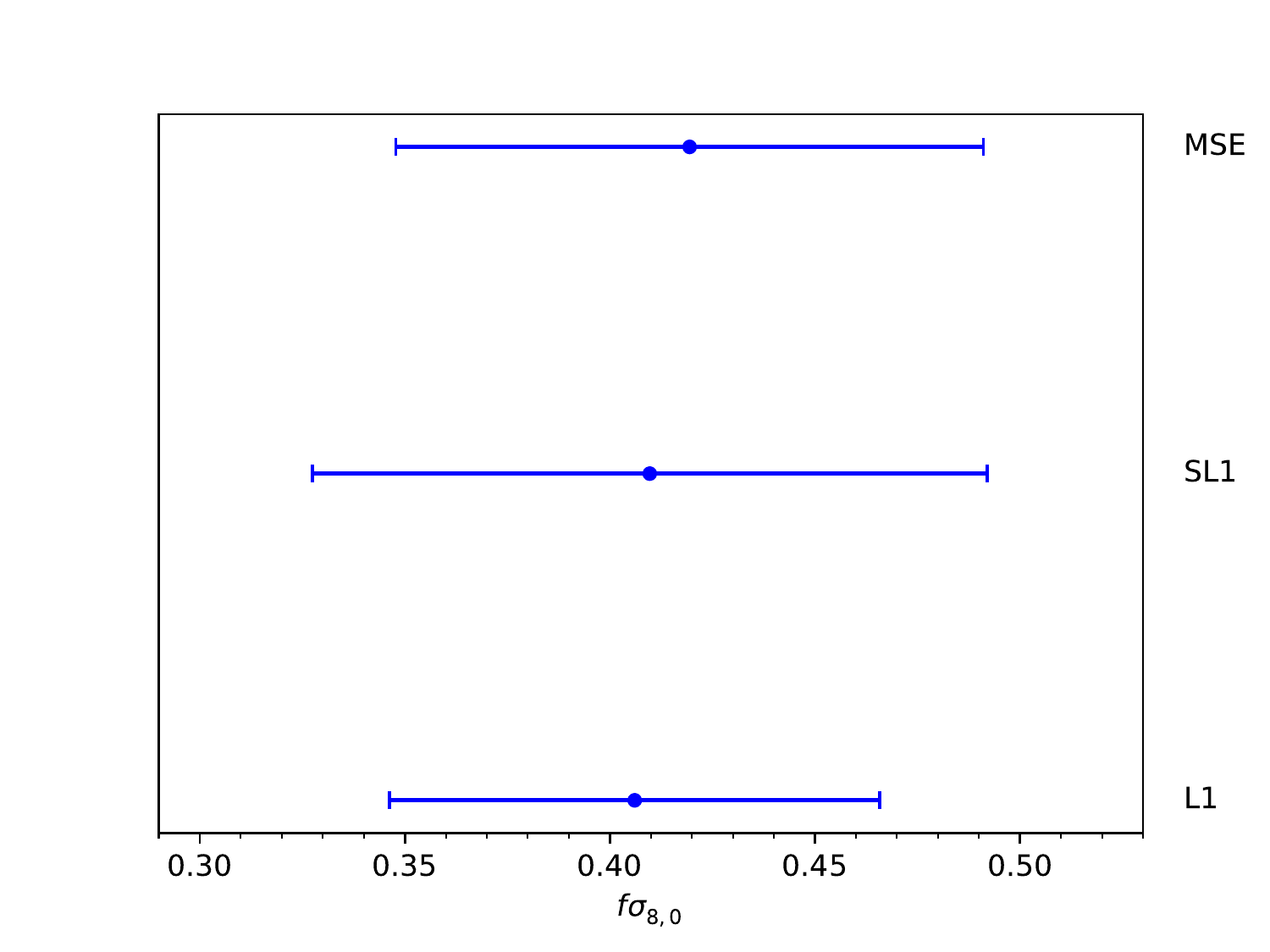}
    \caption{We depict the L1 $f\sigma_8(z)$ ANN reconstructions with different number of layers and neurons in the top-left and top-right panels, respectively. In the bottom-left panel we depict the $f\sigma_8(z)$ ANN reconstructions adapting the L1, SL1 and MSE loss functions. In the latter panels, the corresponding $f\sigma_8(z)$ data points are also included. In the bottom-right panel we illustrate the inferred $1\sigma$ constraint on $f\sigma_{8,0}^{}$ from the $f\sigma_8(z)$ ANN reconstructions as indicated on the vertical axis.}
    \label{fig:fs8_ANN}
\end{figure}

We now generate a mock data set of $f\sigma_8(z)$, for which we also need to take into account the uncertainty of the observational $f\sigma_8(z)$ data points. In the right panel of Fig. \ref{fig:fs8z_data}, we illustrate the errors of the considered $f\sigma_8(z)$ data set as a function of redshift. Similar to the $H(z)$ data, the uncertainties of the $f\sigma_8(z)$ data set tend to increase with the redshift. We will therefore be assuming a first degree polynomial function of redshift \cite{Ma:2010mr,Velasquez-Toribio:2021ufm} for the error of $f\sigma_8(z)$. The mean fitting function is found to be $\sigma_{f\sigma_8}^0(z)=0.07+0.02z$, while the symmetric upper and lower error bands are respectively specified by $\sigma_{f\sigma_8}^+(z)=0.09+0.05z$ and $\sigma_{f\sigma_8}^-(z)=0.05-0.01z$. We further depict these fitting functions in the right panel of Fig. \ref{fig:fs8z_data}, in which one could easily observe that the majority of the data points are included in the area enclosed by the $\sigma_{f\sigma_8}^+(z)$ and $\sigma_{f\sigma_8}^-(z)$ functions (dashed lines). 

We shall now proceed with the random generation of the errors for our $f\sigma_8(z)$ mock data points, in which we assume that the error $\tilde{\sigma}_{f\sigma_8}^{}(z)$ follows the Gaussian distribution $\mathcal{N}(\sigma_{f\sigma_8}^0(z),\,\varepsilon_{f\sigma_8}^{}(z))$, where $\varepsilon_{f\sigma_8}^{}(z)=(\sigma_{f\sigma_8}^+(z)-\sigma_{f\sigma_8}^-(z))/4$, such that $\tilde{\sigma}_{f\sigma_8}^{}(z)$ falls in the area with a probability of 95\%. Therefore, every simulated growth rate data point ${f\sigma_8}_\mathrm{sim}^{}(z_i)$ at redshift $z_i$, is computed via ${f\sigma_8}_\mathrm{sim}^{}(z_i)={f\sigma_8}_\mathrm{fid}^{}(z_i)+\Delta {f\sigma_8}_i$, with the associated uncertainty of $\tilde{\sigma}_{f\sigma_8}^{}(z_i)$, where $\Delta {f\sigma_8}_i$ is determined via $\mathcal{N}(0,\,\tilde{\sigma}_{f\sigma_8}^{}(z_i))$.

Similar to the cosmic expansion ANN analysis, the entire $f\sigma_8(z)$ mock data points were used to train the network. The set--up for the determination of the optimal ANN structure follows from the adopted methodology for the cosmic expansion data. Indeed, the initial learning rate is also set to 0.01 which will decrease with the number of iterations, while the training batch size is set to half of the number of the observational data points. Also, the network is trained after $10^5$ iterations, such that the loss function no longer decreases after this number of iterations, which is clearly illustrated in the right panel of Fig. \ref{fig:fs8_risk_loss}. From the risk parameter, we determined that the network model with one hidden layer is the best performing neural structure. Furthermore, from the normalised risk values $\mathcal{R}$ as depicted in the left panel of Fig.~\ref{fig:fs8_risk_loss}, one could clearly notice that 4096 neurons minimise the risk function, and therefore the latter number of neurons with one hidden layer shall be adopted in our analyses. Furthermore, we shall be using the L1 loss function, since this was characterised by the lowest risk statistic with respect to the MSE and SL1 loss function networks.

\subsection{\label{sec:ANN_fs8z_Omz}ANN \texorpdfstring{$f\sigma_8^{}(z)$}{} reconstructions and \texorpdfstring{$\mathcal{O}m_{f\sigma_8^{}}^{}(z)$}{} null test}

\begin{figure}
    \centering
    \includegraphics[height=0.32\linewidth]{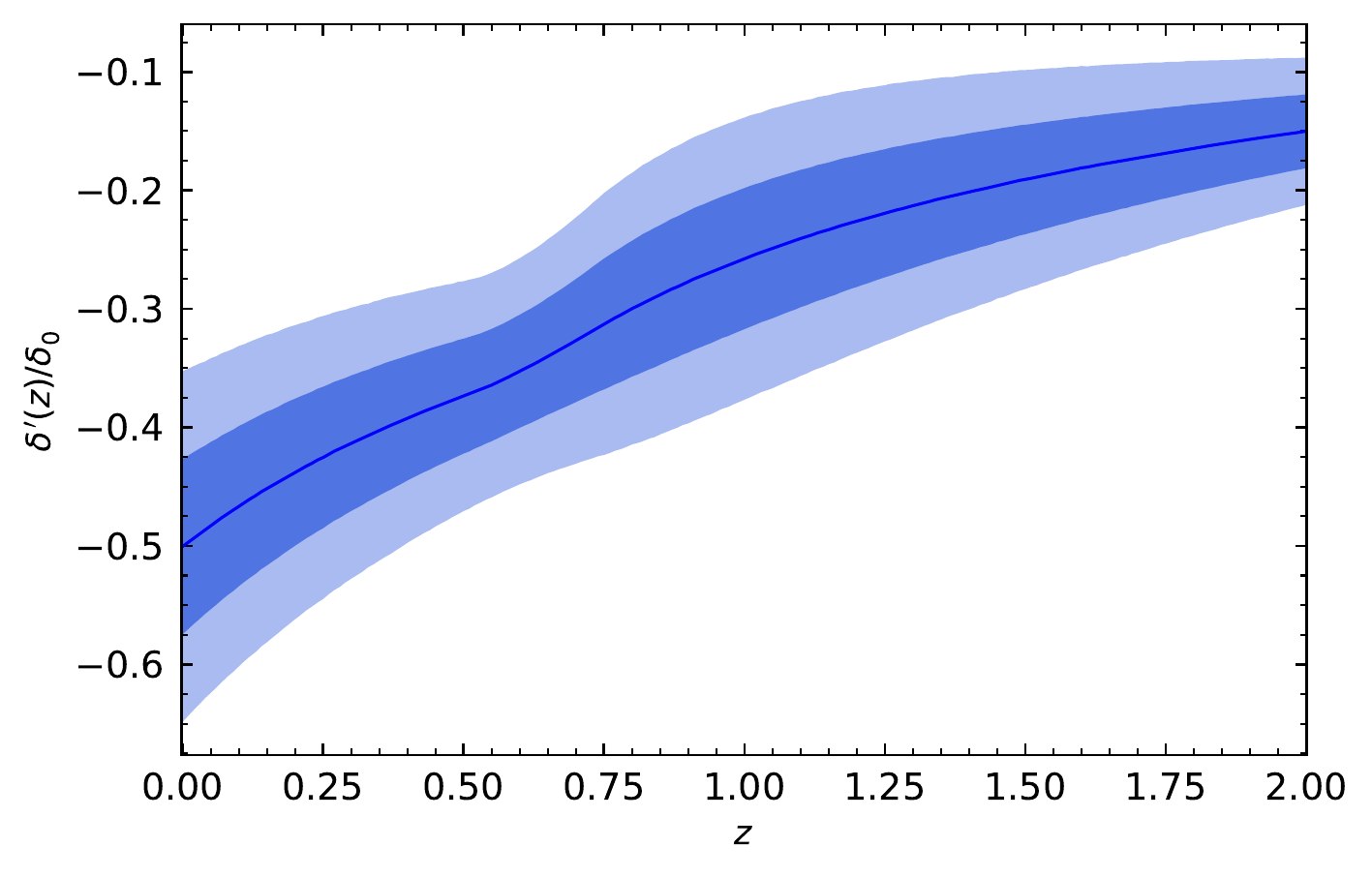}
    \includegraphics[height=0.32\linewidth]{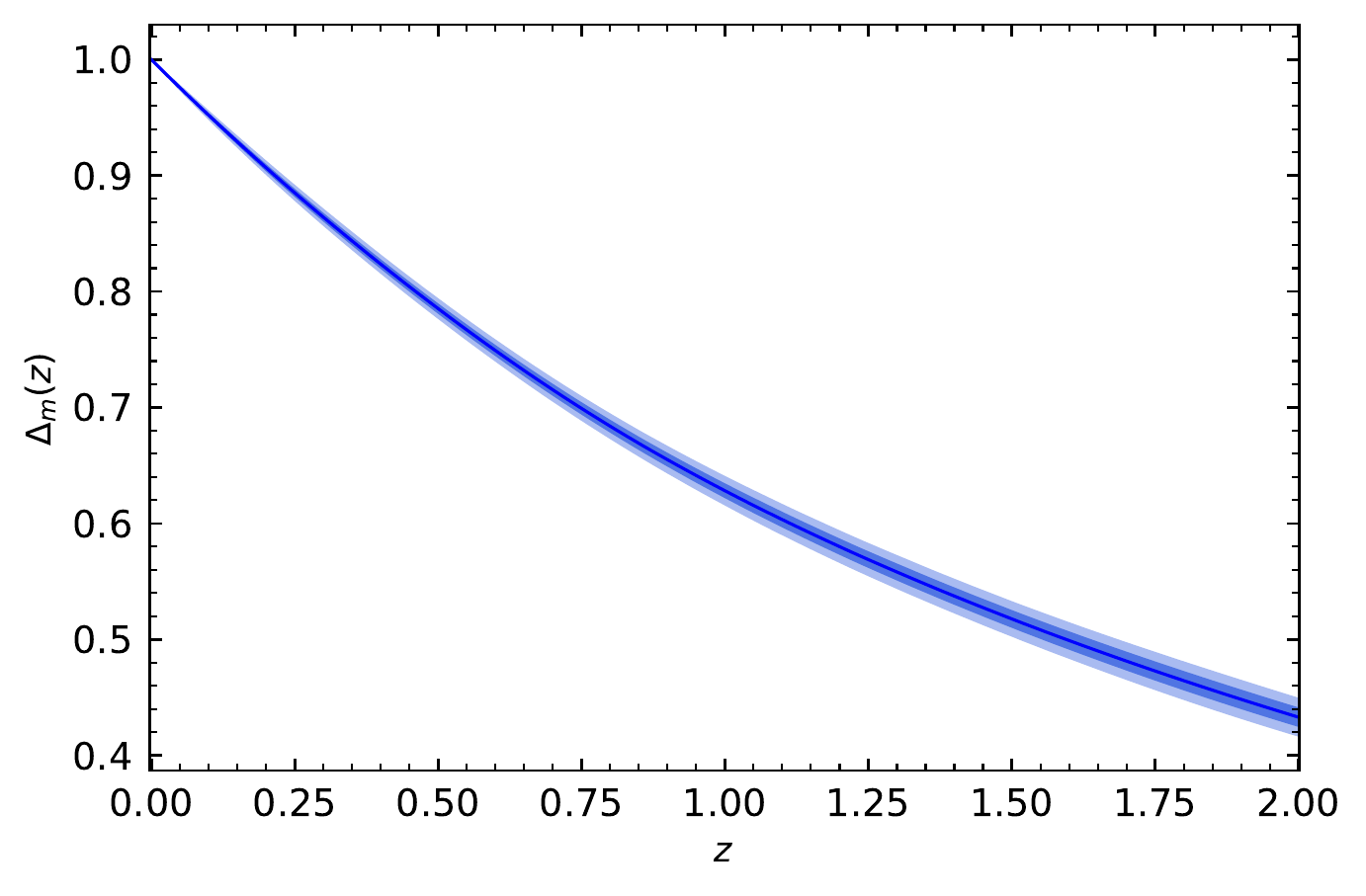}
    \includegraphics[width=0.585\linewidth]{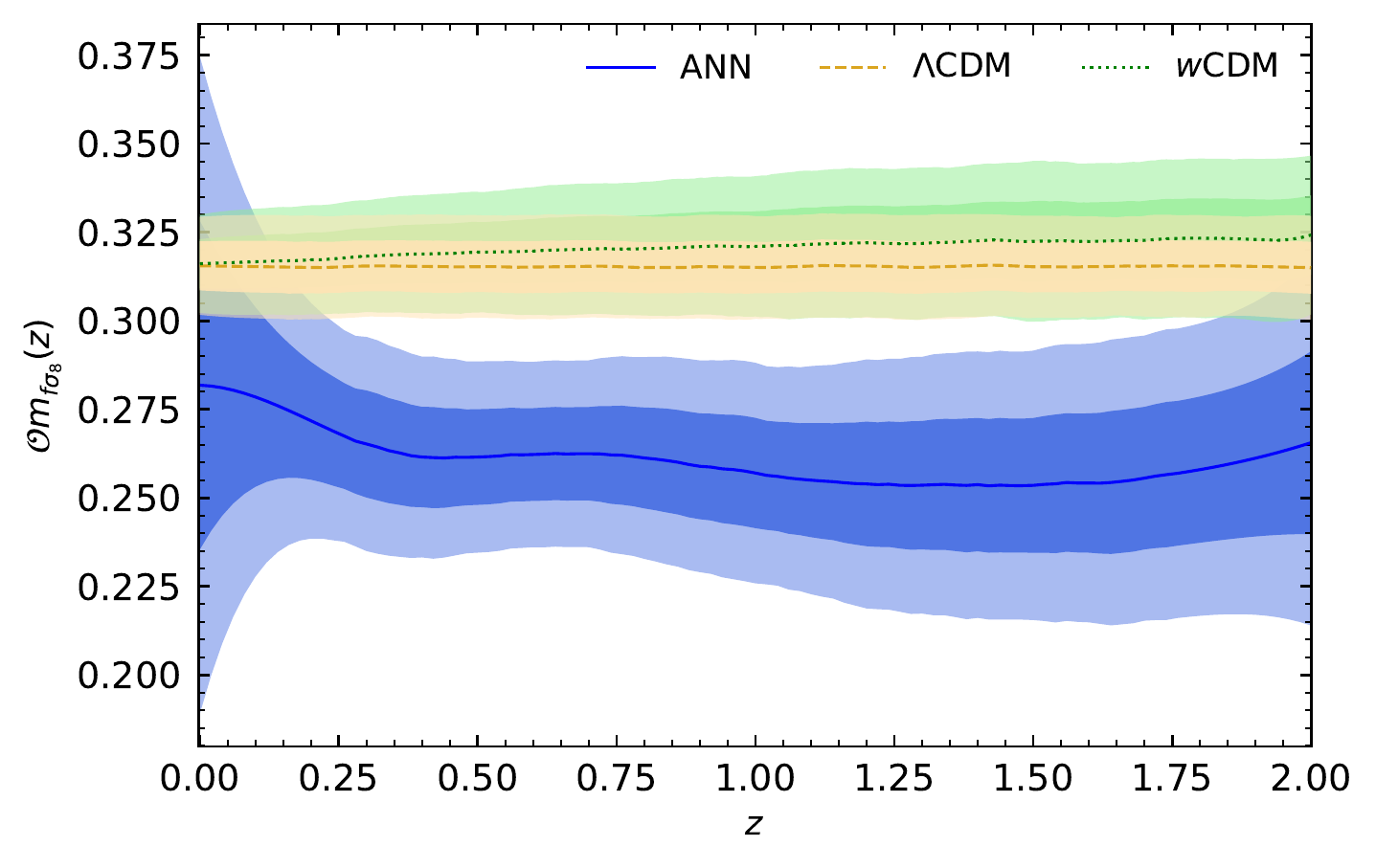}
    \caption{ANN reconstructions of $\delta^\prime(z)/\delta_0$ (top--left) and $\Delta_m(z)$ (top--right) with the L1 loss function. In the bottom panel we illustrate the L1 ANN reconstruction of the $\mathcal{O}m_{f\sigma_8}(z)$ null test, along with the $\Lambda$CDM and $w$CDM predictions when adopting the \textit{Planck} parameter constraints.}
    \label{fig:delta_fs8}
\end{figure}

We now shift to the ANN reconstruction of observational $f\sigma_8(z)$ data with the optimal ANN structure of one hidden layer and 4096 neurons. We illustrate the ANN $f\sigma_8(z)$ reconstructions in the top and bottom-left panels of Fig. \ref{fig:fs8_ANN} along with the observational data points. Similar to the ANN reconstruction of the cosmic expansion data set, we notice that when one considers more layers, this leads to more complex features in the reconstruction which will adversely impact the performance of the reconstruction, as shown in the top--left panel of Fig. \ref{fig:fs8_ANN}. In the top--right panel of the latter figure, one could notice that the number of neurons affects the ANN reconstruction, which was not observed in the $H(z)$ data set. Furthermore, the loss functions also have a mild impact on the ANN reconstruction of the cosmic growth data set, as depicted in the bottom-left panel. We further compare the inferred $1\sigma$ constraint on the value of $f\sigma_{8,0}\equiv f\sigma_8(z=0)$ in the bottom-right panel of Fig. \ref{fig:fs8_ANN}, in which one could observe that the SL1 ANN reconstruction tends to be characterised by a larger uncertainty. One could also notice that the reported GP reconstructions \cite{Pinho:2018unz,Li:2019nux,LeviSaid:2021yat} are in agreement with the derived ANN reconstructions. For comparison purposes, we also reconstruct the $\delta^\prime(z)/\delta_0$ function as illustrated in the top--left panel of Fig. \ref{fig:delta_fs8}, in which this function is also in agreement with the GP reported function \cite{LeviSaid:2021yat}, although slightly more conservative with larger error bars.  

For the growth rate null test, we shall first consider the quantity $\Delta_m(a)$, specified by \cite{Arjona:2021mzf}
\begin{equation}\label{eq:Delta_fs8}
    \Delta_m(a)=\frac{\delta_m(a)}{\delta_m(a_0)}=\frac{a\,{}_{2}F_{1}\left[\frac{1}{3},1;\frac{11}{6};a^3\left(1-\frac{1}{\Omega_{m,0}}\right)\right]}{{}_{2}F_{1}\left[\frac{1}{3},1;\frac{11}{6};\left(1-\frac{1}{\Omega_{m,0}}\right)\right]}\,,
\end{equation}
where we adopted the conventional normalisation of $a_0=1$. The ANN reconstruction of this function is illustrated in the top--right panel of Fig. \ref{fig:delta_fs8}. From this equation we need to express the matter density parameter $\Omega_{m}(z)$ as a function of $f\sigma_8(z)$ and the redshift, i.e. $\mathcal{O}m_{f\sigma_8^{}}^{}(z)$. We shall therefore consider a series expansion on Eq. (\ref{eq:Delta_fs8}) around $\Omega_{m,0}^{-1}=1$ up to the first fifteen terms, up to which the accuracy of this series expansion with respect to the analytical expression was found to be at the sub--percent level. We then apply the Lagrange inversion theorem to invert the series expansion and write the inverse matter density $\Omega_m^{-1}(z)$ as a function of $\Delta_m(a)$, i.e. $\Omega_m^{-1}(z)\equiv\mathcal{O}m_{f\sigma_8^{}}^{-1}(z,\,\Delta_m)$, leading to the $\mathcal{O}m_{f\sigma_8^{}}^{}(z)$ null test which is specified by $\mathcal{O}m_{f\sigma_8^{}}^{}(z)=\frac{1}{\mathcal{O}m_{f\sigma_8^{}}^{-1}(z,\,\Delta_m)}$. 

With this theoretical framework of the cosmic growth null test, from the ANN reconstruction of $f\sigma_8(z)$, we reconstruct the function of $\Delta_m(z)$ via Eq. (\ref{eq:Delta_fs8}) which will then be used to reconstruct $\mathcal{O}m_{f\sigma_8^{}}^{}(z)$. This is considered as a null test of the concordance model of cosmology, since in $\Lambda$CDM $\mathcal{O}m_{f\sigma_8^{}}^{}(z)=\Omega_{m,0}$, hence any deviations from this equality would lead to a data driven departure from the $\Lambda$CDM model. The ANN reconstructed cosmic growth null test is illustrated in Fig. \ref{fig:delta_fs8}, along with the current limits on $\mathcal{O}m_{f\sigma_8^{}}^{}(z)$ within the $\Lambda$CDM and $w$CDM models when adopting the \textit{Planck} model parameter constraints. In case of $w$CDM, the following analytical evolution of the matter density contrast was adopted for the illustration in Fig. \ref{fig:delta_fs8} \cite{Buenobelloso:2011sja}
\begin{equation}
    \delta_m^{w\mathrm{CDM}}(a)=a\,{}_{2}F_{1}\left[-\frac{1}{3w},\frac{1}{2}-\frac{1}{2w};1-\frac{5}{6w};a^{-3w}\left(1-\frac{1}{\Omega_{m,0}}\right)\right]\,,
\end{equation}
with $w=-1.028\pm0.032$ \cite{Aghanim:2018eyx}. From the ANN reconstruction of $\mathcal{O}m_{f\sigma_8^{}}^{}(z)$, it is clear that the \textit{Planck} constraints on $\Lambda$CDM and $w$CDM models are much tighter for all the considered redshift range. Moreover, one could also observe that the ANN reconstructed cosmic growth null test is not found to be in good agreement with the $\Lambda$CDM and $w$CDM predictions, particularly between $0.25\lesssim z\lesssim 1.9$ in which the discrepancy is between $2\sigma$ and $\sim3.5\sigma$. Such a discrepancy supports the already reported tensions \cite{Joudaki:2019pmv,Asgari:2019fkq,Benisty:2020kdt,DiValentino:2020vvd} between the early--time and late--time cosmological data sets probing the large scale structure, although in our case the reported departure from the concordance model of cosmology is data driven and model--independent. 

\section{\label{sec:conc}Conclusions}

The arena of nonparametric reconstruction techniques has drastically grown in recent years due to the increasingly polarizing features of the emergence of cosmological tensions in cosmological data sets. In tandem, this has also been coupled to the introduction of a number of important null tests on concordance cosmology. By and large, the central approach to these reconstruction approaches has relied on GP to various extends which suffers from several critical deficiencies which can be principally summarized by the kernel selection problem, in which different kernels must be surveyed for consistency among each other, and the overfitting of Gaussian uncertainties particularly at low redshifts, which are a focal point of interest in this context.

In this work, we have explored the possibility of using trained ANNs to reconstruct late--time cosmological data where we similarly assume Gaussian uncertainties for comparisons' sake. We did this both for cosmic expansion data and for the growth of large scale structure data. ANNs utilize an unsupervised process for learning meaning that the structure of the ANN is of crucial importance in accurately mimicking input data. For this reason, in both cases, we first worked on designing the ANN structure in Secs.~\ref{sec:hubble_training} and \ref{sec:fs8_training} where the lack of complexity in the data led to a preference for a one layer system with similar numbers of neurons in the network. As discussed in these sections, the mock data was produced in a very simplified manner, a choice which is consistent with the eventual reconstructions that follow. Other approaches led to near identical results in this regard.

The reconstructed $H(z)$ and $f\sigma_8(z)$ profiles are respectively shown in Figs.~\ref{fig:Hz_ANN} and \ref{fig:fs8_ANN} where the one layer systems clearly approximate the data much better than the other options which exhibit more oscillatory behavior. The reconstructions are also largely stable for increasing numbers of neurons in this respect. On this point, the reconstructions both appear consistent across the various loss functions which again points to the resilience of these reconstructed profiles. $H(z)$ is slightly different to $f\sigma_8(z)$ in that there are drastically different priors that have appeared in the literature in recent years while $f\sigma_8(z)$ continues to tolerate a slightly lower tension between literature values. For this reason, we also show how priors on $H_0$ impact the reconstruction of this expansion parameter in Fig.~\ref{fig:Hz_ANN}. Surprisingly, it turns out that the $H(z)$ reconstructions are largely independent of any choice of prior. This means that the ANN gives a more equal weighting to the distributed observation points rather than the initial or low redshift elements.

Beyond the reconstructions of the expansion and growth profiles, we also perform null tests to examine the consistency of the reconstructions against an $\mathcal{O}m(z)$ diagnostic. This diagnostic can help reveal any preference for deviations away from the $\Lambda$CDM concordance model. In Fig.~\ref{fig:Om_z} the diagnostic for the reconstructed $H(z)$ data gives large uncertainties for low values of the redshift which then diminish for larger redshifts. This occurs due to a divergence at $z=0$ and is independent of the choice of prior, as shown in the separate plots. In Fig.~\ref{fig:delta_fs8}, we show this diagnostic for the $f\sigma_8(z)$ reconstructions. We similarly show the region plots for both $\Lambda$CDM and $w$CDM using Planck parameter values. Here, a sizable difference arises between the reconstructed data and its model analogues which may point to the need for the consideration of other null tests.

The ANN approach is drastically different to the GP method where the over-fitting and kernel selection issues have been largely suppressed or eliminated. The source of these properties is in the fact that ANNs have a much larger number of hyperparameters which are optimized. This means that the resulting trained ANN is a much better imitation of the natural process producing the observations as compared with GP. Another important observation of the ANN technique is that its $H(z)$ reconstruction is largely independent of literature priors unlike its GP counterpart which is significantly altered for different priors. One expects these foundational differences to emerge due to the fact that the ANN structure assumes far less than GP in this context, thus giving a more authentic profile for these cosmological parameters.

It would be interesting to probe other data sets and forecast observations for surveys expected to report results over the next few years. It may also be intriguing to explore other null tests of the concordance model which may further expose the impact of reported cosmological tensions. However, ultimately it would be one would want to couple these different cosmological parameters in some model independent way through a process involving an ANN structure. Alternatively, one may also want to use these ANN reconstructions to constrain possible cosmological models such as modified gravity and dark matter models.

\begin{acknowledgments}
The authors would like to acknowledge networking support by the COST Action CA18108 and funding support from Cosmology@MALTA which is supported by the University of Malta. This research has been carried out using computational facilities procured through the European Regional Development Fund, Project No. ERDF-080 ``A supercomputing laboratory for the University of Malta''. The authors would also like to acknowledge funding from ``The Malta Council for Science and Technology'' in project IPAS-2020-007. KFD acknowledges support by the Hellenic Foundation for Research and Innovation (H.F.R.I.) under the “First Call for H.F.R.I. Research Projects to support Faculty members and Researchers and the procurement of high-cost research equipment grant” (Project Number: 2251).
\end{acknowledgments}

\bibliographystyle{JHEP}
\bibliography{references}

\end{document}